\documentclass[manuscript, nonacm]{acmart}

\AtBeginDocument{%
  }

\begin{document}

\title[Navigating Automated Hiring]{Navigating Automated Hiring: Perceptions, Strategy Use, and Outcomes Among Young Job Seekers}

\author{Lena Armstrong}
\email{larmstrong@g.harvard.edu}
\orcid{0000-0001-8306-9270}
\affiliation{%
  \institution{Harvard University}
  \city{Cambridge}
  \state{MA}
  \country{USA}
}

\author{Danaé Metaxa}
\email{metaxa@seas.upenn.edu }
\orcid{0000-0001-9359-6090}
\affiliation{%
  \institution{University of Pennsylvania}
  \city{Philadelphia}
  \state{PA}
  \country{USA}
}

\renewcommand{\shortauthors}{Lena Armstrong and Danaé Metaxa}

\begin{abstract}
  As the use of automated employment decision tools (AEDTs) has rapidly increased in hiring contexts, especially for computing jobs, there is still limited work on applicants' perceptions of these emerging tools and their experiences navigating them. To investigate, we conducted a survey with 448 computer science students (young, current technology job-seekers) about perceptions of the procedural fairness of AEDTs, their willingness to be evaluated by different AEDTs, the strategies they use relating to automation in the hiring process, and their job seeking success. We find that young job seekers' procedural fairness perceptions of and willingness to be evaluated by AEDTs varied with the level of automation involved in the AEDT, the technical nature of the task being evaluated, and their own use of strategies, such as job referrals. Examining the relationship of their strategies with job outcomes, notably, we find that referrals and family household income have significant and positive impacts on hiring success, while more egalitarian strategies (using free online coding assessment practice or adding keywords to resumes) did not. Overall, our work speaks to young job seekers' distrust of automation in hiring contexts, as well as the continued role of social and socioeconomic privilege in job seeking, despite the use of AEDTs that promise to make hiring ``unbiased.'' 
\end{abstract}





\maketitle

\section{Introduction}
\label{sec:intro}

The use of automated employment decision tools (AEDTs) has rapidly increased in hiring contexts, especially for computing jobs. This includes the use of software in reading resumes, scoring coding assessments, and even conducting interviews to assess job applicants among others~\cite{bogen2018help, sanchez2020does}. Job applicants in this changing hiring landscape have developed strategies to navigate these new tools, such as using online coding assessment practice tools or putting their resumes through resume scanners~\cite{armstrong2023navigating}. The automated systems with which job seekers must contend are typically built by third-party companies for employers, and their development largely ignores the perceptions and experiences of the applicants interacting with them. While prior work has identified a lack of auditing, accountability, and transparency in this space~\cite{ajunwa2019auditing, sloane2022silicon, sanchez2020does}, things are beginning to change as policy-makers start to introduce auditing laws for AEDTs, such as New York City's Local Law 144, which mandates employers publish some auditing results for any AEDTs they use in hiring~\cite{locallaw144}. However, meta-audits of the hiring audits published in compliance with Local Law 144 have found significant limitations, including considerable employer discretion about whether a system constitutes an AEDT and is therefore covered by the law, what data is used, what results are actually published in the audits, and where the findings are publicized (making them largely inaccessible to job seekers)~\cite{wright2024null}. As AEDTs and related law and policy proliferate, it is imperative to understand job applicants' own perceptions of AEDTs, the strategies they use to navigate this environment, and how strategies relate to their job market outcomes.

Understanding applicants' beliefs about and practices surrounding the use of automation in hiring is important for understanding the impact of these tools, improving such technologies when they are used, and restricting their use when inappropriate. Investigating human perceptions is imperative to building trustworthy, fair, and explainable AI in a range of domains~\cite{woodruff2018qualitative, lee2018understanding, glikson2020human}. Prior work has identified that one factor contributing to people's negative perceptions of algorithmic decision-making is a feeling of dehumanization caused by being algorithmically evaluated~\cite{lee2018understanding}. Another is distrust in the system, which can be caused by the perception that the system is not fair~\cite{woodruff2018qualitative, lee2018understanding}. Using a procedural fairness framework, which investigates fairness through the processes by which decisions are made as opposed to the outcomes~\cite{lee2019procedural}, we build on on prior work in this space to investigate how applicants perceive and navigate different forms of AI-based automated hiring technologies, as well as the impact, if any, of such strategies on hiring outcomes. 

While many automated hiring companies claim to reduce bias in hiring, prior work has shown that many automated hiring systems can be as biased as human-decision makers~\cite{caliskan2017semantics}, and that some even violate U.S. and U.K. anti-discrimination laws~\cite{ajunwa2019paradox, sanchez2020does}. For computing students, who are often navigating the job market for the first time, biases introduced by automated hiring methods may be especially impactful, affecting their applications' success, but also their sense of belonging in the field, overall anxieties, and even their ability to enter the workforce at all~\cite{behroozi2019hiring, kasinidou2021agree}. Some initial work on applicant awareness of AEDTs has found that applicants have experience with a variety of AEDTs, even if they do not know they are interacting with an AEDT, and report using strategies to navigate automated hiring, such as getting referrals, modifying resumes for automated scanners, and using online coding assessment practice materials in response to them~\cite{armstrong2023navigating}. While previous work explores bias in automated hiring or applicant perceptions of automation, this is a still-growing area, with more research needed on the intersection of applicants' perceptions of various forms of automation, experience with automated hiring strategies, and job outcomes. 

To better understand young job seekers' perceptions and experiences with automated hiring, we pose the following research questions: 

% RQs:
\begin{enumerate} 
    \item[\textbf{RQ1}] How do young job seekers' perceptions of AEDTs' procedural fairness and their own willingness to be evaluated by different AEDTs change as the level of automation and type of evaluation change?
    \item[\textbf{RQ2}] Do perceptions of AEDTs' procedural fairness and willingness to be evaluated by AEDTs change based on applicants' use of strategies to navigate automation in the hiring process?
    \item[\textbf{RQ3}] How do awareness about AEDTs, strategies to navigate automation, and demographic attributes (like gender, race, and income) relate to job market outcomes?
\end{enumerate}

% Method: 
To answer these questions, college-aged computer science students (relevant for their status as young, current job seekers) answered a two-part survey. In the first, they were given descriptions of different scenarios describing a hypothetical hiring process, each of which included an \textit{evaluation type} (coding assessment, resume review, or interview) and \textit{automation level} (ranging from human-only to AI-only review). For each scenario, they were asked to assess the fairness of each hiring process \textit{(procedural fairness measure)} and whether they themselves would want to be evaluated that way \textit{(willingness measure)}; this data was analyzed to evaluate their perceptions of and attitudes towards automated hiring algorithms. In the second part, participants provided information on whether they had experienced or heard about AEDTs (coding assessments, resume scanners, etc.) and what strategies they used relating to these practices (coding assessment practice, adding keyword to their resumes, etc.). We analyzed this data to evaluate whether participants' perceptions of automated hiring processes were related to their strategy use, and whether those strategies had an impact on job market success (whether they reported receiving job offers).

We found that participants across the board rated the use of AI as unfair compared to human-only evaluation, and that as the level of automation increased, procedural fairness and willingness ratings decreased. Additionally, participants found the use of AI more inappropriate in less technical evaluations. For example, fairness and willingness ratings for the use of AI in coding assessments were higher than for behavioral interviews. Notably, examining the relationship between participants' strategy use and demographics with job outcomes, we found that only two attributes were significantly positively associated with job market success: the degree to which participants used referrals and their family's household income. 

Overall, our findings point to widespread dislike among young job seekers towards AEDTs, as well as the continued role (not `fixed' by automation) of social and socioeconomic privilege in job seeking. Our work makes contributions to CSCW by providing new insight into young job-seekers' perspectives on AEDTs and the strategies they use to succeed in an increasingly automated hiring market. We conclude by discussing implications for auditing, policy, and use of AEDTs by employers in the future.

\section{Background}
\label{sec:background}

We review previous research related to automated employment decision tools (AEDTs) as a whole, as well as work on how AEDTs impact various stakeholders' perceptions and applicant career-seeking. 

\subsection{Automated Employment Decision Tools}
Throughout the hiring process, recruiters increasingly rely on AEDTs to identify open positions and manage applicants, check and score resumes, evaluate technical skills, and even automate video interviews~\cite{bogen2018help, sanchez2020does}. The development of such technologies is an active area of research. For instance, prior work has built resume scanning tools using natural language processing~\cite{sanyal2017resume, harsha2022automated, bharadwaj2022resume} and by matching keywords in job descriptions to resumes~\cite{dhende2018candidate, satheesh2020resume} to assess candidates. Additionally, researchers have investigated how to use AI to screen applicants using automated video interviews where applicants record a video of themselves answering questions in a fixed amount of time. Researchers have then used AI to assess personality traits and evaluate job ``fit,'' through a variety of verbal and nonverbal behaviors, such as word choice, fluency, pronunciation, inflection, facial expressions, posture, and eye movements ~\cite{chen2016automated, naim2016automated, hickman2022automated}. However, other scholars have critiqued these methods, expressing concerns about their ethics and validity~\cite{roemmich2023values, rhea2022resume}. 

Such critiques are often connected to a research angle focused on algorithmic fairness. Prior research has shown that AI can learn implicit bias similar to humans, measured by word and image associations~\cite{caliskan2017semantics}. In the context of hiring, researchers have shown that AEDTs can not only be biased, but that they may do so in violation of anti-discrimination laws~\cite{ajunwa2019auditing, sanchez2020does, raghavan2020mitigating}. In one highly-publicized example, Amazon stopped using their resume reviewing AEDT after machine-learning specialists discovered it scored resumes lower when they mentioned the word ``woman'' or the names of historically women's colleges~\cite{dastin2018amazon}. Work has found gender bias in other systems as well, for example showing that some automated systems rate women's resumes lower than men's even when controlling for similar levels of experience and job-relevant traits~\cite{parasurama2022gendered, chen2018investigating}. In addition to gender bias, previous auditing work has found biases in such systems along the lines of race~\cite{ajunwa2019auditing}, disability status~\cite{buyl2022tackling}, and age~\cite{farber2017factors} among other factors. 

As new legislation like New York City's Local Law 144, which mandates employers publish the results of a bias audit for any AEDT they use, is enacted, there is a growing need to understand the perceptions and experiences of job seekers themselves, who otherwise have limited input in the hiring processes to which they are subjected. Recent work has found that current legislation is missing key definitions and auditing standards, does not prevent companies from using biased AEDTs, and gives employers an inappropriately high degree of discretion on who conducts the audits and what results are published from them~\cite{groves2024auditing, wright2024null}. Given this context, there is a crucial need for applicants' own perspectives to inform audits and policy. Our study contributes to this literature by providing insights into applicants' perceptions about AEDTs' procedural fairness, their willingness to be evaluated by AEDTs, and the formal and informal strategies they use to navigate the automated hiring landscape.

\subsection{Perceptions of Algorithmic Fairness}

It is imperative to understand not only bias within automated systems, but also people's perceptions of algorithmic fairness, which directly impact trust~\cite{woodruff2018qualitative, lee2018understanding}. One approach to understanding user perceptions of fairness in algorithmic decision-making is through \textit{procedural fairness}, which refers to an evaluation of the process by which a decision is made as opposed to its outcome~\cite{lee2019procedural, schoeffer2021appropriate}. Procedural fairness stands in contrast to other kinds of fairness perceptions of automated systems, including those focusing on outcomes (distributive fairness), interactions (interpersonal fairness), and  explanations (informational fairness)~\cite{colquitt2012organizational, schoeffer2021appropriate}. When comparing types of fairness perceptions, procedural fairness has been shown to be a robust predictor of overall fairness evaluations and heavily relied on by users to judge the fairness of AI systems~\cite{thibaut1975procedural, morse2021ends, glikson2020human}. Components that impact procedural fairness perceptions include the perceived consistency, competency, and benevolence of the decision-maker~\cite{lee2019procedural, morse2021ends, leventhal1980should, tyler2003procedural}. Additionally, how much voice individuals have over the decision process, such as the ability to provide feedback or overturn a flawed decision, can impact procedural fairness and system trust~\cite{lee2019procedural, morse2021ends, lind1990voice, folger1977distributive, jeung2023correct}. In this work, we measure people's perceptions of AEDTs' procedural fairness, building on existing literature in the space by investigating its relationship to participants' willingness to be evaluated by such AEDTs.

Numerous factors have been shown to impact people's perceptions of algorithmic fairness, such as the task being automated~\cite{lee2018understanding, zhang2022examining}, whether the system's decision is in the user's favor~\cite{wang2020factors}, and the amount of human oversight~\cite{lee2017algorithmic, gonzalez2022allying}. For tasks that require more human skills, such as hiring or work performance evaluations, people view algorithmic decision-makers as less fair, effective, and trustworthy than mechanical skills, like work assignment or scheduling~\cite{lee2018understanding}. Previous qualitative work on computing students' perceptions of automated hiring found that applicants felt human recruiters could assess skills that automated processes could not, and that this limited their ability to demonstrate the full extent of their skills when being automatically evaluated~\cite{armstrong2023navigating}. In general, entirely algorithmic decisions are perceived as less fair than decisions made by humans~\cite{lee2017algorithmic, gonzalez2022allying}. Human decision-making may also lead applicants to report more positive experiences with hiring due to the potential for social recognition, applicants' sense of being acknowledged and appreciated by another person~\cite{lee2018understanding, armstrong2023navigating}. Despite the current automated hiring landscape, applicants often attribute success in the hiring process to social recognition and connections~\cite{armstrong2023navigating, chua2021playing}. However, there is currently limited work on applicants' perceptions of fairness and their desire to be evaluated by automated versus non-automated systems as it exists in relation to other attributes like social connections.

\subsection{Career-Seeking Amid Automation}
Young job seekers' experiences with an increasingly automated landscape are important for their future success; insufficient support and negative experiences at this stage can impact career-seeking and transitions from school into the workforce~\cite{bock2013women, giannakos2017understanding, jones1986socialization, begel2008novice}. Perceptions and experiences in automated hiring may influence applicants' desire to apply for certain positions and sense of belonging in the field. Past work found that perceptions of algorithmic fairness impacted user trust with implications for tool use~\cite{woodruff2018qualitative}. 
When trust is lost, qualified applicants may be dissuaded from applying through automated hiring process with limited transparency \cite{van2021job}. This may be especially true of applicants coming from underrepresented backgrounds---the same applicants who are often most negatively impacted by biased AEDTs---whose decisions about whether to stay in the field are heavily influenced by environmental factors, presence of feedback, and presentation of information~\cite{bock2013women, giannakos2017understanding, metaxa2018gender}.
Prior work has found a disconnect between employers' and applicants' hiring process priorities~\cite{friedrich1993primary}, but as human resources professionals begin to utilize AEDTs, there is some research to suggest that even some HR workers have concerns about data accuracy and ceding control to automated decision-making systems~\cite{li2021algorithmic}. 

Given job seekers' concerns, they must develop new strategies to navigate automation. Qualitative research interviewing job-seekers has revealed the importance of referrals in circumventing automation and securing job offers, and that they often attribute their hiring successes to knowing how to ``play the hiring game''~\cite{armstrong2023navigating, chua2021playing}. Students from different socioeconomic backgrounds have reported different experiences with hiring processes even when they have access to the same resources, with applicants from working and middle-class backgrounds feeling like they need to earn social credit before relying on social connections~\cite{chua2021playing}. Consequently, automated hiring experiences and knowledge of strategies may not only impact applicants desire to stay in the computing field~\cite{olson2014opportunities}, but also their ability to receive job offers. Further work is needed to assess applicants' experiences and perceptions of fairness in relation to socioeconomic status and resource access in the context of automated hiring processes. We expand on prior qualitative work by providing a quantitative survey investigation, examining how referrals and other strategies impact people's perceptions of procedural fairness, desire to be evaluated by AEDTs, and job-seeking success. 

\section{Method}
\label{sec:methods}

We conducted a survey of computer science students to investigate their perceptions of various automated hiring technologies and how their strategy use in this landscape impacts these perceptions as well as job success.

\subsection{Participants \& Recruitment}
After an initial power analysis for small to moderate effect size ($0.25-0.30$) suggested we should recruit at least 170-250 participants, we surveyed 525 computer science undergraduate and graduate students in two phases. Across both phases of recruitment, the recruitment message described the research team as being interested in understanding fairness in automated hiring and hoping to share what they learn to inform better automated hiring practices and career service support. Participants were informed that the survey would take 15 minutes and ask them to respond to scenarios about different hiring practices and describe their recruitment experiences. The University of Pennsylvania's Institutional Review Board approved our initial study, along with modifications when recruitment was expanded to other schools. Additionally, we pre-registered our hypotheses and analyses on Open Science Framework~\cite{armstrong2023computing}.

\subsubsection{Recruitment Phase 1. }In the first phase, 294 participants were recruited from the University of Pennsylvania, a private university in Philadelphia. These students were enrolled in an upper-level artificial intelligence course, and were given extra course credit for completing our survey. This was initially intended to be the full scope of the study, but we ultimately elected to delay analysis and instead diversify our participant sample outside of the University of Pennsylvania, expanding it to two additional universities in the same city. 

\subsubsection{Recruitment Phase 2. }Lacking connection to a specific course with an extra credit research incentive (as we were able to do in the first phase of recruitment), we instead reached out to professors at two other institutions who distributed our survey to their students with a monetary incentive. We offered to enter participants into a gift card raffle with limited success. In pursuit of obtaining enough participants at the other universities to balance the nearly-300 from the University of Pennsylvania, we used several additional strategies. These included recruiting through flyers on those campuses, emailing computing clubs' leadership and faculty colleagues at the two institutions, posting on computing employment-related social media forums, contacting administrators related to school-wide surveys and job boards, and even tabling on those campuses near their computer science buildings. We also emailed all previous participants encouraging them to refer a friend with an incentive of another entry in the gift card raffle. We found more success when we reached out again to professors to distribute the survey to their students with another non-monetary incentive, individualized feedback on students' resume. All strategies considered, we were able to get an additional 231 responses. 

\subsubsection{Data Cleaning. }Our 525 survey responses were reduced to 448 after removing unfinished surveys ($n=31$), respondents who were not computing students at the University of Pennsylvania, Temple University, or Drexel University ($n=16$), those who failed attention checks ($n=28$), and duplicate email addresses ($n=2$). After data cleaning, we analyzed 284 responses from the University of Pennsylvania, 153 from Temple University and 11 from Drexel University. Note that in the following analyses (Section \ref{sec:results}) we do not analyze for differences by university. Due to the different recruiting strategies used, students at the University of Pennsylvania were predominantly upper-level, while those at Temple University and Drexel University were more junior, so we found it inappropriate to compare by this attribute. We also collected information about whether students were applying for jobs and whether their job-searching process was complete or still underway.

\begin{table}[t]
\begin{tabular}{lllll}
\cline{1-2} \cline{4-5}
\textbf{Gender}               & \textbf{Count} &  &\textbf{ Race \& Ethnicity}               & \textbf{Count} \\ \cline{1-2} \cline{4-5} 
Agender              & 3     &  & American Indian or Alaskan Native      & 1     \\
Genderqueer          & 2     &  & Asian American or Asian         & 251   \\
Man                  & 238   &  & Black or African-American       & 46    \\
Non-binary           & 6     &  & Hispanic or Latinx              & 18    \\
Woman                & 167   &  & Middle Eastern or North African & 8     \\
Prefer Not to Answer & 36    &  & Pacific Islander                & 1    \\ \cline{1-2}
                     &       &  & White or Caucasian              & 82    \\
                     &       &  & Prefer Not to Answer            & 40    \\ \cline{4-5} 
\end{tabular}
\vspace*{3mm}
\caption{\label{tab:gender/race} Participant demographics for gender and race \& ethnicity}
\end{table}

\subsubsection{Demographics. }
\begin{figure}[t]
    \centering
    \includegraphics[width = 0.7\textwidth]{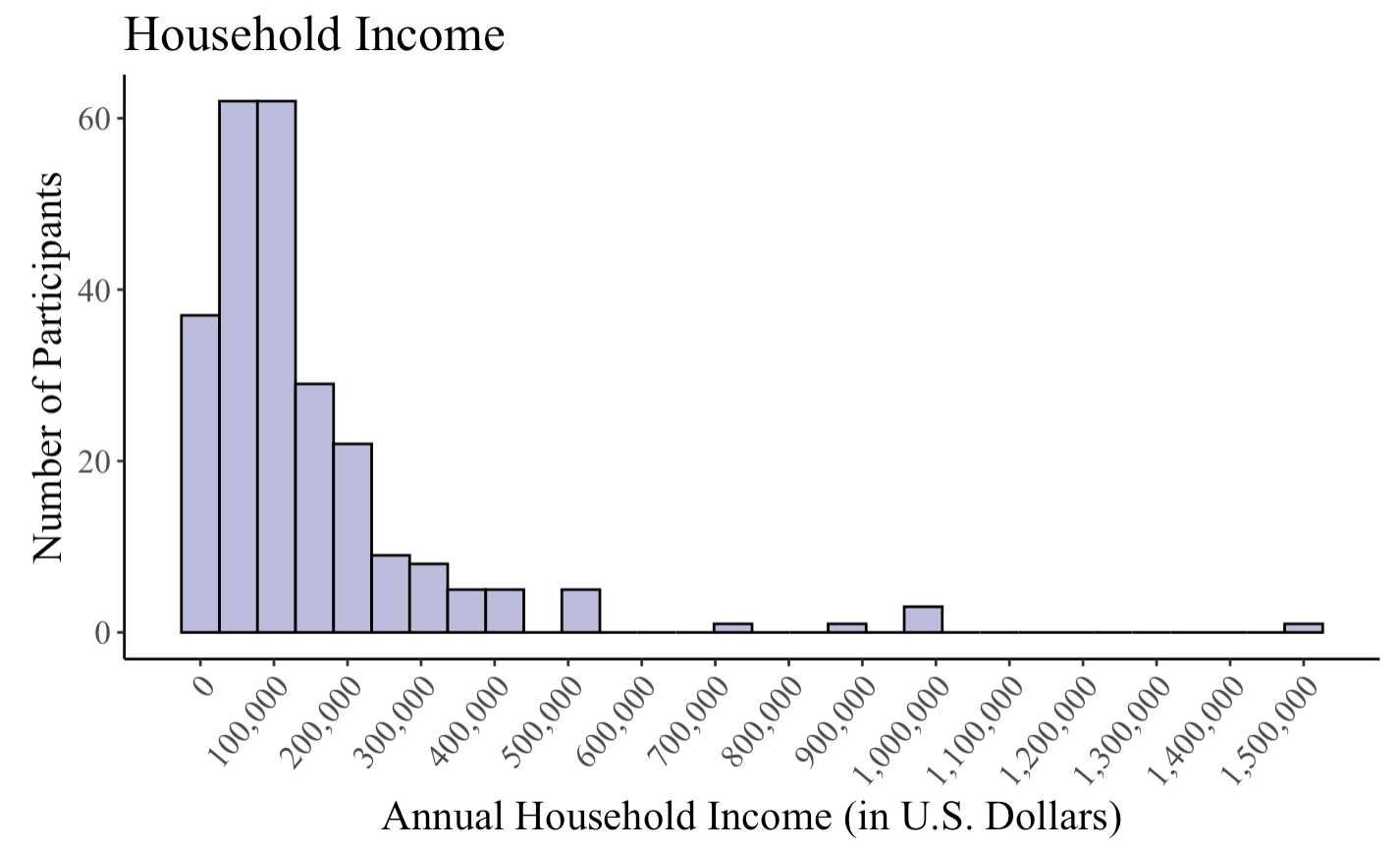}
    \caption{Participant self-reported household income from \$0 to \$1,500,000, with a median of \$100,000.}
    \label{fig:income}
    \Description[Histogram of participant self-reported household]{Histogram of participant self-reported household income from \$0 to \$1,500,000, with a median of \$100,000.}
\end{figure}

At the end of the survey, participants were asked to describe their gender and racial/ethnic identities (summarized in Table ~\ref{tab:gender/race}). Demographic information questions were created based on inclusive survey guidance~\cite{fernandez2016more} and allowed participants to optionally choose zero, one, or multiple categories that best described them. 250 participants opted to share their family's annual household income, ranging from \$0 to \$1.5 million with a median of \$100,000 (see Figure ~\ref{fig:income}). 

\subsection{Survey}
The survey contained two main parts: participant evaluations of 15 hiring scenarios (Part 1), and questions about participants' own experience with automated hiring processes (Part 2), in addition to demographic questions. For the full survey, see Appendix \ref{app:survey}.

\subsubsection{Perceptions}
The first part of the survey assessed participants' perceptions of procedural fairness in and willingness to be evaluated by 15 different hiring scenarios. These were made up of three \textbf{evaluation types} (technical coding assessment, resume reviewing, and behavioral interviews) and five different \textbf{automation levels} (human-only reviewing, human-AI reviewing equally, human reviewers for AI rejections, human reviewers for AI acceptances, AI-only reviewing). 

For example, the five scenarios pertaining to an \textit{online coding assessment} evaluation type were: 

\begin{enumerate}
  \item \textit{Human-only reviewers:} An applicant submits a sample of code, which is reviewed by a recruitment team who determines whether the applicant advances to the next phase.
  \item \textit{Human-AI reviewing equally:} An applicant is given an online coding assessment, which is evaluated by an algorithm. If the applicant reaches a certain score on the autograder, the applicant advances to the next phase. All algorithmic decisions are reviewed by a recruitment team.
  \item \textit{Human reviewers for AI rejections:} An applicant is given an online coding assessment, which is evaluated by an algorithm. If the algorithm rejects the applicant, the decision is reviewed by a recruitment team.
  \item \textit{Human reviewers for AI acceptances:} An applicant is given an online coding assessment, which is evaluated by an algorithm. If the algorithm advances the applicant, the decision is reviewed by a recruitment team.
  \item \textit{AI-only reviewing:} An applicant is given an online coding assessment, which is evaluated by an algorithm that determines whether an applicant advances to the next phase.
\end{enumerate}

% True positives are when the organization hires a candidate who is a job match. True negatives are when the organization rejects a candidate who is not a match. False positives are when the algorithm accepts the candidate, but the candidate wasn’t a good fit. Lastly, false positives are when the algorithm rejects a candidate, but the candidate was qualified. 

The evaluation types were chosen based on prior work interviewing computing students about the types of automation they encountered in the hiring process~\cite{armstrong2023navigating} and on the most prevalent automated employment decision tools (AEDTs) for the screening of candidates~\cite{sanchez2020does, bogen2018help}. We chose to focus on technical coding assessments, resume reviewing, and behavioral interviews where applicants receive limited transparency and feedback. Prior work has shown that young job seekers, especially in computing, must navigate automated hiring process more frequently in earlier hiring stages before progressing to human interviews and that some are even hired through entirely automated processes~\cite{armstrong2023navigating, sanchez2020does, bogen2018help}. The different automation levels were chosen based on prior work on perceptions of automated decision-making~\cite{zhang2022examining, lee2018understanding,binns2018s}. We chose to include scenarios for human review of both AI rejections and acceptances, since previous work suggests that both conditions are important to applicants~\cite{friedrich1993primary, roulin2014interviewers, morse2021ends, de2021use}. 

Participants responded to each scenario on a five-point Likert scale, rating both \textbf{procedural fairness}:``this hiring process seems fair'' (1: Strongly disagree -- 5: Strongly agree) and \textbf{willingness}: ``I want to be evaluated this way'' (1: Strongly disagree -- 5: Strongly agree). These options were determined based on best practices with Likert scale use in learning about perceptions ~\cite{jebb2021review, joshi2015likert} and understanding fairness perceptions in automated systems~\cite{lee2019procedural, schoeffer2021appropriate, morse2021ends}. This practice of soliciting and aggregating responses to understand procedural fairness perceptions is consistent with prior CSCW and related work on perceptions of AI systems~\cite{lee2019procedural, lee2018understanding, zhang2022examining, gonzalez2022allying}.

\subsubsection{Awareness of and Experiences with Automated Hiring}
The second part of the survey asked participants to answer questions about their own awareness of AI systems in hiring. They were asked about whether they had experienced or heard about several specific AEDTs (online coding assessments, automated resume readers, and automated video interviews) along with descriptions of those tools. Participants chose from four possible responses (had experienced it, had heard of it but not experienced it, had not heard of or experienced it, or did not know). They also answered whether ``I know how my data was used in the hiring process'' and ``I received feedback from automated hiring algorithms'' on a five-point Likert scale (1: Strongly disagree --- 5: Strongly agree).

They were also asked yes/no questions about their own strategies for navigating the job market, including whether they had modified resumes with keywords, used tools like LeetCode to practice for online coding assessments, gotten referrals, or had family and friends who worked at companies. These strategies were chosen based on prior work interviewing students about the strategies they used to navigate AEDTs~\cite{armstrong2023navigating} as well as past work that resume scanners rely on parsing with natural language processing~\cite{sanyal2017resume, harsha2022automated, bharadwaj2022resume} and keywords in job descriptions~\cite{dhende2018candidate, rhea2022resume, satheesh2020resume} to assess candidate matches. We also asked whether they were currently applying to jobs (300 of our 448 final participants were), how many companies they had applied to, and whether they had received any job offers. Applicants who reported they were not applying for jobs were excluded from analysis about strategy use and job outcomes.

At the end of the survey, participants optionally provided demographic information about their gender, race, and annual household income, factors we hypothesized could play a role in their job market success based on prior work on perceptions of automated decision-making~\cite{lee2018understanding, wang2020factors}. 

\section{Findings}
\label{sec:results}

We proposed these hypotheses and statistically analyzed survey responses to answer each research question:

\begin{enumerate}
\item[RQ1:] How do young job seekers' perceptions of AEDTs' procedural fairness and their own willingness to be evaluated by different AEDT hiring processes change as the level of automation and type of evaluation change?
\item[\textit{H1:}] \textit{As the level of automation in hiring scenarios increases, and as the technical nature of the evaluation type increases, perceptions of fairness will decrease.}
\item[ ] Prior work has shown that perceptions of fairness increase for more mechanical tasks~\cite{lee2018understanding, zhang2022examining} and when a human decision maker is included~\cite{gonzalez2022allying}, which we hypothesized will also hold in our study, in regards to participants' reported procedural fairness evaluations of different AEDTs and willingness to be evaluated by them.
\end{enumerate}

\begin{enumerate}
\item[RQ2:] Do perceptions of AEDTs' procedural fairness and willingness to be evaluated by AEDTs change based on applicants' use of strategies to be successful in the hiring process?
\item[\textit{H2:}] \textit{Participants with greater use of strategies will report lower perceptions of fairness.}
\item[ ] We suspected that applicants who have more resource access and greater awareness of AEDTs might view the system as less meritocratic and more game-able, and thus distrust their fairness ~\cite{leclercq2020gamification, leutner2023game}.
\end{enumerate}

\begin{enumerate}
 \item[RQ3:] How do awareness about AEDTs, strategies to navigate automation, and demographic attributes (like gender, race, and income) relate to job market outcomes?
\item[\textit{H3:}] \textit{Participants with higher AEDT-awareness and higher reported strategy use will obtain more job offers.}
\item[ ] We hypothesized that applicants with more experience and awareness of automated systems would be better at navigating automated hiring systems and therefore see more positive job outcomes based on prior work showing applicants attribute their successes to awareness of how to play the ``game'' in hiring~\cite{armstrong2023navigating, chua2021playing}.
\end{enumerate}

We found evidence to support the first and second hypotheses. Participants largely did not consider automated processes fair or want to be evaluated by them, and they rated greater automation (less human involvement) and automation of less technical evaluations (i.e., interviews as opposed to coding assessments) lower across on both measures.
When looking at the relationship between perceptions of fairness and strategy use, the use of strategies like referrals, as well as awareness of some AEDTs was associated with lower fairness ratings. 
We found partial evidence supporting the third hypothesis; use of referrals was the only strategy associated with job market success. The other major predictor of job success was family household income.

\subsection{Participants' Perceptions of the Hiring Process}

We begin by answering RQ1: how do young job seekers' perceptions of AEDT procedural fairness and willingness change as a result of automation level and evaluation type?

In the first part of the survey, participants rated the procedural fairness and their own willingness to be evaluated by different hiring scenarios on a 5-point Likert scale. Each scenario, as described in Section~\ref{sec:methods}, had an \textbf{evaluation type} (online coding assessment, resume, or interview) and \textbf{automation level} (human-only, human-AI, AI with human review of rejections, AI with human review of acceptances, and AI-only). Participants rated both (1) how \textit{fair} they found the hiring process in the scenario, and also (2) their \textit{willingness} to be evaluated that way themselves. We found that participants' ratings of procedural fairness were generally similar to their own willingness to be evaluated by these methods, but that the  willingness measure was usually slightly lower (confirmed with a paired t-test comparing all ratings of (1) to ratings of (2) ($t(6718) = 23.27$, $p < 0.001$)).

\subsubsection{Evaluation Type} 
\begin{figure}
\centering
  \includegraphics[width = 0.74\textwidth]{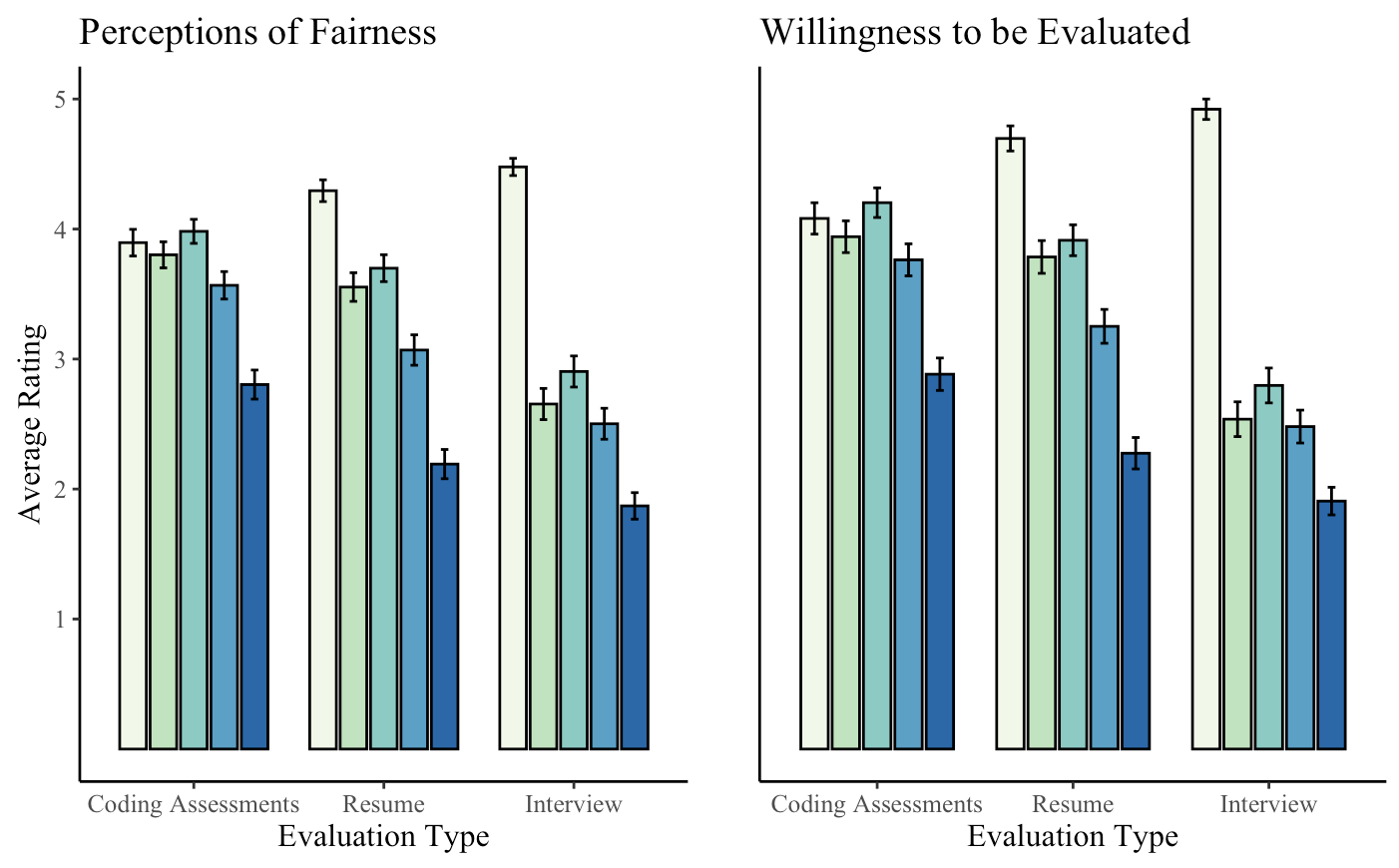}
  \includegraphics[width = 0.24\textwidth]{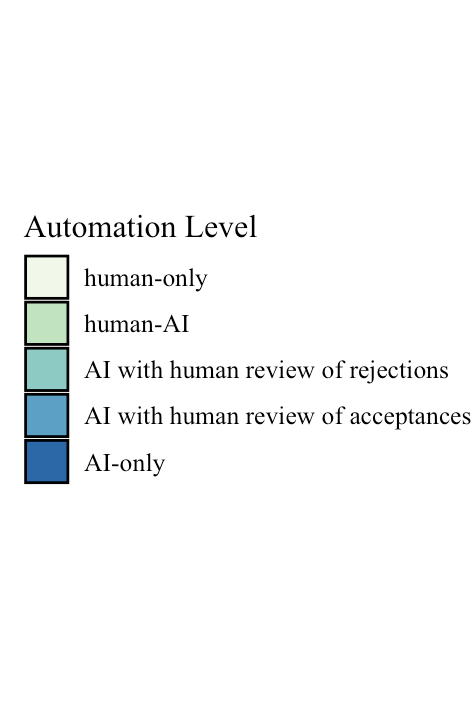}
  \Description[Bar chart comparing fairness ratings for a range of evaluation types and automation levels.]{Bar chart where human-only evaluation is preferred over AI, especially for interview assessments.}
  \caption{Participants' \textit{procedural fairness} and \textit{willingness} perceptions for different evaluation types and levels of automation were rated on a Likert scale (1: process seems very unfair / unwilling to be evaluated this way --- 5: seems very fair / willing to be evaluated this way). Notably, both measures performed similarly with participants. Responses reflected that some human involvement was always preferred, and AI-only review was least-preferred across evaluation types. However, there was a greater tolerance for AI involvement in more technical evaluations (i.e., in coding assessments compared to interviews). Bars represent 95\% confidence intervals.} 
  \label{fig:scenarios}
\end{figure}

Comparing participant perceptions of fairness in the three evaluation types (online coding assessment, resume, and interview), we found that participants rated the use of automation (all decision-makers except ``human-only'') as less fair in less technical evaluation types than more technical ones. Automation was perceived as most fair for a coding assessment evaluation and least fair for a behavioral interview evaluation (see Figure ~\ref{fig:scenarios}). We conducted an ANOVA for the effects of evaluation type ($F(2) = 238.15$, $p < 0.001$) and automation level ($F(4) = 518.93$, $p < 0.001$) on procedural fairness perceptions, finding significant effects for each, as well as their interaction ($F(8) = 51.19$, $p < 0.001$). We also conducted an ANOVA for the effects of evaluation type ($F(2) = 267.07$, $p < 0.001$), automation level ($F(4) = 521.17$, $p < 0.001$), and their interaction ($F(8) = 68.37$, $p < 0.001$) on participants' own willingness to be evaluated in this way, with consistent results of the significant effects.
We did not find any significant effects of race, gender, or household income on procedural fairness or willingness perceptions.

Several participants verbalized the idea that less technical evaluation types should not be automated in their responses to an open-ended question in which we asked for their thoughts on this part of the survey. One participant explained, ``[A] resume is much harder to quantify via an algorithm than piece of code.'' Another explained, ``resumes are much more nuanced than coding assessments, as applicants can have different strengths and weaknesses which can qualify them differently.''
The idea that the use of automation in hiring processes leads them to be ``gamed'' and therefore less fair caused participants to bristle.
One wrote, ``Much more than resumes and coding assignments, automating the actual interview removes the human aspect of jobs and turns it into a game. Not only could interviewees cheat the system, exploit bugs, and cause chaos, but turning human interaction into a soulless computer guessing game makes the interview much less about how much an employer likes a candidate but how much the candidate can understand the algorithm judging it.''

\subsubsection{Automation Level} 
Regarding automation levels (human-only, human-AI, AI with human review of rejections, AI with human review of acceptances, and AI-only), fairness and willingness ratings were higher for lower levels of automation across evaluation types (see Figure ~\ref{fig:scenarios}). Participants found the scenarios more fair when there was a human in the loop, and the AI-only conditions always ranked lowest. Summarizing this perspective in an AI-only interview scenario, one participant wrote, ``This is the worst so far, human interaction can't be replaced with a machine. Personally, I've taken an automated interview and it felt wrong.''

Within the human-AI combination decision-makers, participants preferred a human evaluator to review AI decisions for rejections (checking for false negatives) over acceptances (checking for false positives) (Figure ~\ref{fig:scenarios}). A one-sided paired t-test confirmed this finding as statistically significant for both the fairness ($M= 0.48$, $CI= 0.43$, $t(1343)=15.92$, $p < 0.001$) and willingness ($M= 0.43$, $CI= 0.37$, $t(1343)=13.37$, $p < 0.001$) measures.

This aligned with our hypotheses about job applicants' priorities; as one participant expressed, ``an algorithm rejecting a deserving applicant is a lot worse than advancing an undeserving applicant.''
Overall, participants perceived human evaluators as a guardrail against AI, which they perceived as potentially more biased. One participant summarized: ``I think that there should be some aspect of human review, whether or not the applicant passes doesn't really matter, I think there should just be some sort of fail-safe.''

\begin{table}
\begin{tabular}{lr} \hline
\textbf{Strategy Use} & \textbf{Percentage} \\ \hline                    
    \hspace{2mm}Used online resources              & 81\% \\ 
    \hspace{2mm}Practiced for online coding assessment              & 69\% \\ 
    \hspace{2mm}Talked with people who had recently applied              & 63\% \\ 
    \hspace{2mm}Used referrals              & 57\% \\ 
    \hspace{2mm}Used application materials and descriptions              & 57\% \\ 
    \hspace{2mm}Had friends who worked at companies                      & 51\% \\
    \hspace{2mm}Mass-applied to companies (more than 20)                 & 50\% \\ 
    \hspace{2mm}Used career services through university                  & 45\% \\
    \hspace{2mm}Modified resumes using keywords from the job description & 42\% \\
    \hspace{2mm}Connected with recruiter through company                 & 34\% \\         
    \hspace{2mm}Put resume through a resume scanner                      & 26\% \\       
    \hspace{2mm}Had family members who worked at companies               & 19\% \\              
    \hspace{2mm}Connected with recruiter outside of company              & 12\% \\ \hline
\end{tabular}
\caption{\label{tab:strategies} Most participants used a variety of strategies to navigate automated hiring processes, especially  online resources, practice for online coding assessments, and talking with others who had recently applied.}
\end{table}

\subsection{Strategy Use}

In these results, we answer research questions RQ2 (Do perceptions of AEDTs' fairness and willingness to be evaluated by AEDTs change based on applicants' use of strategies to be successful in the hiring process?) and RQ3 (How do awareness about AEDTs, strategies to navigate automation, and demographic attributes like gender, race, and income relate to job market outcomes?).

\subsubsection{Descriptive Results}
\begin{figure}
\centering
\begin{minipage}{.45\textwidth}
  \centering
  \includegraphics[width=1\linewidth]{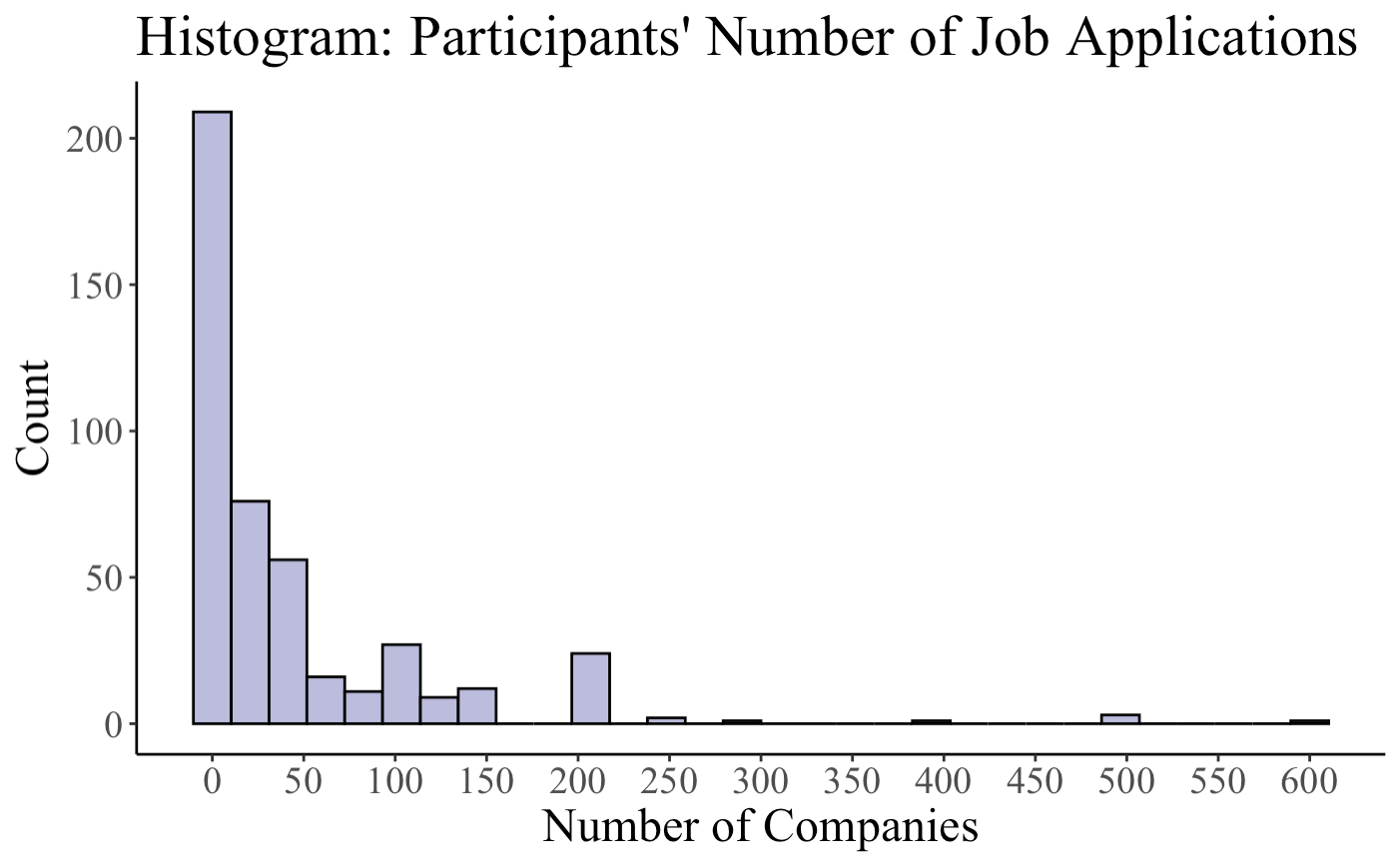}
  \Description[Histogram of number of jobs participants applied to]{Histogram with long tail showing most participants applied to <15 jobs but many applied to 50 or more}
  \captionof{figure}{Participants reported applying to 0 to 600 jobs, with 50\% applying to more than 20.}
  \label{fig:numCompanies}
\end{minipage}%
\hspace{0.05\textwidth}
\begin{minipage}{.45\textwidth}
  \centering
  \includegraphics[width=1\linewidth]{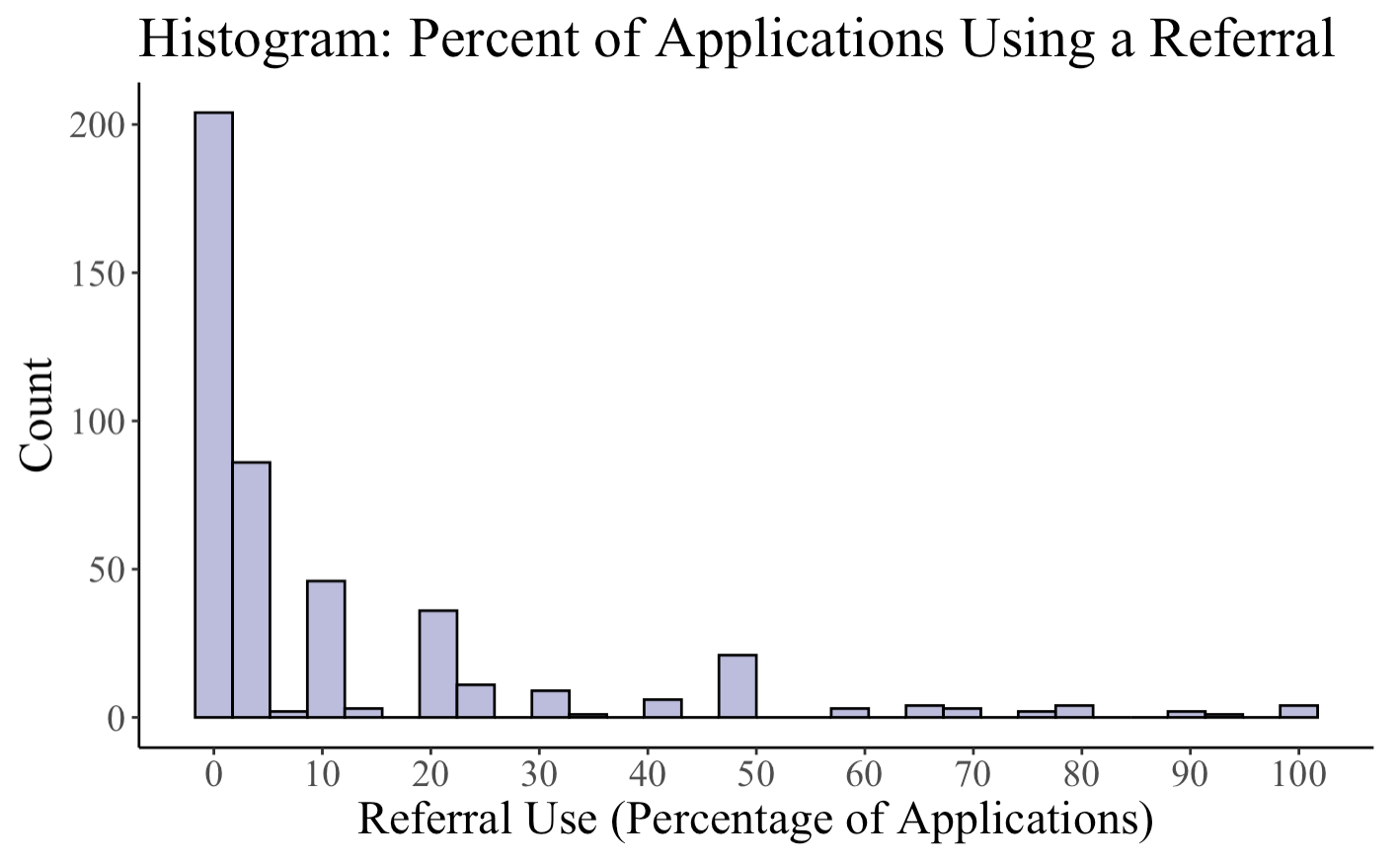}
  \Description[Histogram of percent of applications that used referrals]{Histogram long tail shows hat most participants used <5 referrals but many had 20 or more.}
  \captionof{figure}{Participants used referrals in anywhere from 0\% to 100\% of applications, with over half of participants having at least one referral.}
  \label{fig:referral}
\end{minipage}
\end{figure}

Participants reported differences in strategy use (e.g., online coding assessment practice or modifying resumes for scanners), visualized in Table~\ref{tab:strategies}. While many applicants added keywords from the job description to their resume (42\%), fewer applicants had put their resume through a resume scanner to see how it was processed (26\%). A more commonly-used strategy than either was the use of LeetCode or another coding assessment practice tool (69\%). 
Mass-applying, which we define as applying to more than 20 jobs, was another strategy exactly half of participants reported using. 33\% of participants applied to more than 50 jobs. Overall, participants' number of applications ranged from zero to 600 ($mean = 45$, $median = 17$). Number of company applications are visualized in Figure~\ref{fig:numCompanies}. When applying, 57\% reported receiving at least one referral; applicants reported receiving referrals for an average of 12\% of their job applications, though this was skewed by some on the much higher end ($mean = 12\%$, $median = 3\%$) (Figure \ref{fig:referral}). 

Participants reported learning about the hiring process through many different channels, including online resources (81\%), others who had recently applied (63\%), application materials and descriptions (57\%), friends who worked at companies (51\%), career services through their universities (45\%), company recruiters (34\%), family members who worked at companies (19\%), and recruiters outside of companies (12\%). 
In terms of awareness of these systems, most applicants knew AEDTs existed: 95\% had heard of online coding assessments, 88\% had for automated resume readers, and 87\% for automated video interviews. Most also reported personal experience with these systems: 63\% of participants reported having experience with online coding assessments, 53\% for automated video interviews, 47\% for resume readers, and 38\% for applicant tracking systems. We found less familiarity with automated resume readers and applicant tracking systems, perhaps because these are often not explicitly mentioned as automated in the application process~\cite{schumann2020we, wilson2021building}. Additionally, many participants reported not know how their data was used in the hiring process (77\%) and not receiving feedback throughout the process (83\%).

 \subsubsection{Strategy Use and Participant Perceptions}
Next, we compare participants' perceptions of procedural fairness and their willingness to be evaluated by AEDTs, as these relate to the various strategies participants used. (For a complete list of these strategies, see Table~\ref{tab:strategies}). 
Using a linear model predicting participants' average perceived fairness and willingness ratings from awareness of and use of different strategies (as well as participant demographics), we found that participants who reported \textit{using referrals} (Fairness: $F(200) = -2.478$, $p = 0.014$; Willingness: $F(200) = -2.312$, $p = 0.022$), and who were \textit{aware of online coding assessment tools} like LeetCode (Fairness: $F(200) = -2.349$, $p = 0.020$; Willingness: $F(200) = -2.356$, $p = 0.019$) reported lower perceived fairness and lower evaluation willingness, while \textit{awareness of automated video interviews} was actually associated with higher fairness and willingness scores (Fairness: $F(200) = 2.113$, $p = 0.036$; Willingness: $F(426) = 2.608$, $p = 0.010$).

Two other attributes (participants who reported using online resources in their application process and believed they knew how their data were used) had inconsistent effects on fairness and willingness. The remaining strategies did not  show significant effects on either fairness or willingness, which included modifying resumes with keywords, changing layouts for resume scanners, testing resumes with scanners, practicing for online coding assessments with tools like LeetCode, mass-applying to companies, using career services, or access to contacts within companies. We also did not find significant effects due to any demographic variables, including race, gender, or household income. For the full statistical test results see Appendix Tables~\ref{tab:fairStats} and ~\ref{tab:evalStats}.

\subsubsection{Strategy Use and Job Outcomes}

Finally, we investigated the effect of awareness of AEDTs and strategy use on job outcomes. We defined job success as receiving one or more job offers, and fit a regression to predict job success from strategy use, awareness of AEDTs, gender, race, and income.  We found little evidence to suggest many measures of strategy use (i.e., modifying resumes with keywords, checking resumes with scanners, practicing LeetCode, mass-applying to companies, or talking with recent applicants) impact job outcomes. While using at least one referral was not significant, having a higher percentage of jobs applied to with a referral did have significant effects on hiring outcomes ($F(128)=2.063$, $p < 0.041$). 

The other major factor we identified that related to job outcome success was family income. Even after controlling for strategy use, participants who reported higher family income had more success securing employment ($F(128)=2.530$, $p = 0.013$) (Figure~\ref{fig:jobOffers}). For the full test results see Appendix Table ~\ref{tab:jobStats}.
Several mechanisms might explain this effect. For instance, students from more affluent backgrounds may have more internship experience due to their ability to accept unpaid positions, or their family connections to industry may carry more weight. As one participant succinctly described, ``My father [gave] me a hiring QR code which I used to apply for the job and then was hired after a meeting.''

\begin{figure}
    \centering
    \includegraphics[width = 0.9\textwidth]{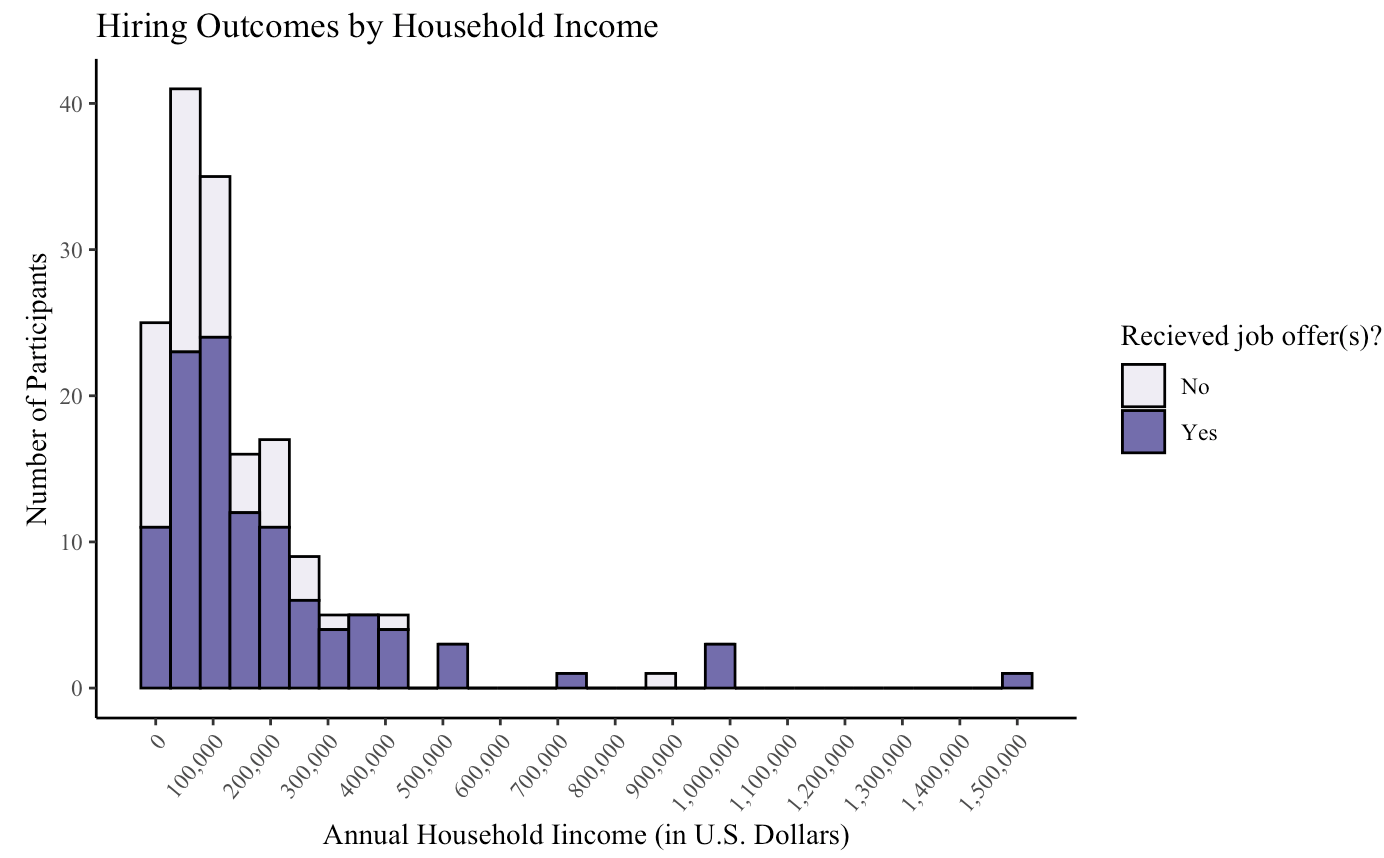}
    \Description[Stacked bar chart of hiring outcome by annual household income]{Stacked bar chart showing that most applicants without job offers came from families making under 200k per year; rates of 1+ job offers skew slightly higher.}
    \caption{Participants reporting higher annual household incomes were also more likely to receive one or more job offers.}
    \label{fig:jobOffers}
\end{figure}

\section{Discussion}
\label{sec:discussion}

We discuss our findings in four parts: the first, about the potential for AEDTs and strategy use, such as referrals to circumvent automation to exacerbate social inequity; next, the implications this has for auditing such systems; third, the role of AEDTs (now and in the future) in holding employers accountable for problematic practices; and finally, directions for future AEDT policies that center workers' needs and experiences. 

\subsection{Exacerbating Social Inequity}
A major finding from this work was the role of individual attributes like socioeconomic status and strategy use, such as referrals in  job market outcomes. In the context of automated employment decision tools (AEDTs), this calls attention to the continued potential for more applicants to gain an advantage by using referrals and other methods to circumvent automated processes altogether. This is a major risk of automation --- that it will result in inequity based not just on the automated evaluations themselves, but in who is subject to them. This will be an important area for future research to continue studying. 

Our findings on socioeconomic differences in hiring outcomes extend current literature that employers factor class-based differences into hiring processes~\cite{adnin2022hiring}. We saw participants in our study achieve greater job success based on annual household income, despite not finding any significant differences in strategy use across all demographic lines, including race, gender, and income. One possible explanation for this phenomenon --- in which financially-privileged applicants report the same knowledge of hiring processes and even strategy use, but still see success at higher rates --- has been discussed in a qualitative investigation into how class differences shape computer science students' experiences and tactics in hiring~\cite{chua2021playing}. That research found that applicants across socioeconomic backgrounds used social connections to secure job offers, had access to connections at companies, displayed similar knowledge of employers' expectations, and faced similar emotional burdens. However, crucially, upper-middle-class participants in that study were more at ease crossing social hierarchy lines than their working- and middle-class counterparts~\cite{chua2021playing}. This manifested in differences in use of specific strategies, including cold-emailing professionals and setting up informal conversations about job opportunities, which ultimately led to referrals that circumvented official hiring pipelines. 
Working- and middle-class job seekers only relied on their closest connections and felt uncomfortable reaching outside that circle for fear of burdening others, whereas upper- and upper-middle-class job seekers felt comfortable using the same strategies to a much greater degree. Perhaps our data's indication of income as important for job market success, while strategy use and awareness of systems had insignificant impact, could be explained by magnitude: differences in people's scale of connections and willingness to use them. 

The impact of social privilege was striking in our findings, suggesting that current hiring practices exacerbate social inequality. Rather than finding that applicant success was driven by factors within their control, we instead found that applicants with greater access to referrals, who were thereby able to evade more automation than their peers, such as resume screeners, and those coming from higher socioeconomic backgrounds, were the ones who received job offers at higher rates. In contrast, more egalitarian strategies available to a wider range of applicants, like free online assessment practice, were not associated with higher levels of job success. Procedural fairness perceptions and willingness to be evaluated by AEDTs were lower among those who received more referrals, which may be explained by those privileged applicants who are able to obtain referrals both knowing about this alternative route to a job offer, and also knowing that many of their peers do not have access to it. Hiring processes that circumvent automation based on personal contacts, such as referrals and informal networks, violate key social justice dimensions, such as \textit{distribution} where the benefits and burdens of AEDTs are not equitably shared, and \textit{enablement} where all people do not have the ability to reach their full potential~\cite{dombrowski2016social}. In contrast to AEDTs that claim to be ``bias-free'' or ``bias-mitigating,'' our results suggest that automated hiring processes continue to perpetuate dominant social inequalities in hiring outcomes, and, newly, in what groups are required to interact with automated hiring tools.

\subsection{Implications for AEDT Auditing}

Speaking to differing perspectives on whether using connections is appropriate, our results also showed that applicants who received referrals had lower ratings of procedural fairness in hiring processes across evaluation types and automation levels. We saw this effect despite the potential for outcome favorability bias~\cite{wang2020factors} (applicants might bias towards thinking the strategies they use are \textit{more} fair). This might be a result of referral-using applicants seeing the system as more gameable, negatively impacting their perceptions of its fairness~\cite{armstrong2023navigating, chua2021playing, leclercq2020gamification}. Meanwhile, applicants without referrals may be unaware that this process exists, or that it can be used to circumvent large parts of the standard application pipeline. Of course, the system is --- and has always been --- gameable in this way. The use of referrals in technology jobs predates any AEDTs. In an ideal world, AEDT use would neither continue nor exacerbate previous human-system biases. Instead, we should re-evaluate current recruiting practices and insist on changes to these systems that provide improvements over prior processes. 

There are important implications for auditing practices towards achieving this outcome. Given the centrality of processes outside any AEDT, algorithm audits cannot stop at analyzing the algorithmic system alone. Instead, they should be done (as other research has recently argued~\cite{lam2023STA}), in the full sociotechnical context in which these tools are used. When applicants circumvent resume screening through referrals or other methods to jump to securing interviews, they gain an advantage that is not captured when auditing the AEDT in isolation. Instead, audits should extend to the entire hiring pipeline at companies using AEDTs (not to mention those which do not). These audits, as we also expressed above, must take into consideration a range of user attributes. Current hiring auditing laws and mandates tend to look at only gender and race~\cite{locallaw144}, but there may be additional and intersectional factors that impact experiences with AEDTs, including income, sexual orientation, age, and disability. 

\subsection{Accountability}
As they currently stand, AEDTs provide employers with two significant benefits: increased efficiency and accountability-avoidance. The first is straightforward; AEDTs allow employers to scale their hiring process and screen many more applicants with smaller HR departments. The latter is significantly more pernicious. As other scholars have also described, the use of AEDTs shifts accountability away from employers and onto the third-party systems they contract~\cite{sanchez2020does, ajunwa2019paradox, wilson2021building}. 

In traditional hiring practices with human decision-makers, the responsibility (or perhaps blame) for unfair hiring decisions could be traced to an individual or team. In contrast, AEDTs allow employers to claim ignorance and instead point the blame at AEDT developers~\cite{wright2024null, groves2024auditing}.  Additionally, across the board, participants reported not knowing how their data was used and not receiving feedback from prospective employers throughout the hiring process. These issues requires swift response from policymakers, such as that seen in New York City, whose recent Local Law 144 (LL144) mandates that employers using AEDTs publicly release a ``bias audit'' document regarding the tools they use~\cite{locallaw144}. 

And yet, there are limitations with such legislation. Most notably, employers are responsible for finding and paying the third-party auditors and publishing the audit results (beyond a few minimal required statistics), giving them out-sized control over these audits' findings~\cite{wright2024null, groves2024auditing}. Moreover, the audits themselves are not auditable! Details on what analysis was conducted on what data, and with what precise results, is not mandated by LL144. Additionally, there may be referrals and informal hiring practices that allow applicants to skip automated hiring processes altogether. Only examining AEDTs ignores the range of potential biases in the hiring pipeline as a whole. Legislation should consider more sociotechnical auditing approaches, looking beyond tool performance to the context in which these tools are used~\cite{lam2023STA}. As it stands, this kind of legislation runs the risk of perpetuating problematic practices and helping employers avoid true accountability. 

Instead of providing avenues for employers to show off their ``bias-free'' practices in a brief, vague document, we need policies aimed at convincingly demonstrating that workers are being treated fairly and lawfully, as well as avenues for recourse and worker-targeted resources for navigating AEDTs when there is (or is suspected to be) foul play. This brings us to our final discussion point, about developing policies that benefit workers, not just employers. 

\subsection{Worker-Centric Policies}
Our findings show that both referral use and the presence of automation breed distrust and perceptions of unfairness in applicants. These, along with the limitations of current legislation outlined above, speak to the need for more human-centered tools and policies. Specifically, in the workplace context, we name this angle \textit{worker-centric}. 

Current practices are completely divorced from workers' desires and their best interests. These are not unreasonable stakeholders --- consistent with prior work~\cite{lee2018understanding, zhang2022examining, gonzalez2022allying}, our participants were quite accepting of automated decisions for more technical scenarios, and relatively tolerant of it when guaranteed a human review of all automated rejection. As it stands, however, since employers are willing to tolerate some false negatives (qualified candidates who do not advance in their process) as opposed to false positives (unqualified candidates who do advance), they and third-party companies have designed AEDTs with the goal of minimizing false positives~\cite{friedrich1993primary}, in direct opposition to job seekers' preferences as studied in this paper and others~~\cite{morse2021ends, roulin2014interviewers}. 

Auditing policies like those focusing on ``equity'' in AEDT scoring or selection rates do not address these important dimensions of fairness during the intermediate steps of the process~\cite{locallaw144}. Legal mandates could help ensure applicants are treated fairly, and simultaneously increase their trust in these processes. In doing so, they would require employers to share in some part of the cost of AEDT use, instead of only relishing the benefits at the expense of job seekers. 

Outside the job market itself, another worker-centered effort could be undertaken by career services groups at universities  to create venues for young job seekers to meet mentors and alumni to bridge socioeconomic divides and work to demystify hidden curricula~\cite{nakai2023uncovering}.

\subsection{Limitations and Future Work}
We note a few key limitations that also provide opportunities for future study in this direction. First, our results are limited in their generalizability by the participant pool we surveyed. This means that our sample may not be reflective of all computing students (those enrolled in other types of institutions or in other geographical regions, in particular), or of all job seekers (especially those who are more senior in the workforce). We did control for whether applicants were finished with their job search; however, since participants took the survey in the fall,  some participants may have received later job offers, which we were not able to capture, introducing inaccuracy into our measure of hiring outcome success. We also did not collect age data from participants, which may also be an important factor in hiring outcomes. We encourage future research to investigate AEDT awareness, perceptions, and experiences with more granularity and in a wider range of settings with regard to location, career stage, age, and other attributes.

Second, our survey questions did not include the gamut of all AEDTs. We selected the specific evaluation methods and applicant strategies in our survey based on prior literature~\cite{armstrong2023navigating}, but may have left out important ones. Moreover, in a space changing as quickly as this one, the systems and strategies used, or people's perceptions about them, may change considerably in the coming years. We encourage future work to continue deepening our understandings of the topics we discussed, and to continue to broaden the research space as needed. 

Finally, while participants in our study reported trusting human evaluation more than AI-involved evaluation processes, we find it important to note that this does not mean human-only processes are free of issues. Indeed AEDT biases have been shown to mirror existing human biases~\cite{caliskan2017semantics}, and also sometimes stem from biases in their training data, which are human-made decisions~\cite{ajunwa2019auditing}. It is possible that our participants inflated their reported judgements of human evaluators in response to the idea of AI review posed in other questions. Future work should continue to study the reality of applicants' experiences navigating human review as well as AEDTs. 

\section{Conclusion}

In this paper, we explore young job seekers' perceptions and strategy use in the face of the increasing prevalence of automated employment decision tools (AEDTs) in workplace hiring practices. To do so, we survey several hundred computer science students from three universities in Philadelphia, asking about their perceptions of procedural fairness and willingness to be evaluated by different levels of automation (ranging from human-only to AI-only review) for different evaluation types (coding assessments, resume reviews, and interviews). We find that students perceive AEDTs as much less fair than human review and have a lower willingness to be evaluated by AI decision-makers for less-technical evaluation types (resume review and interviews). While they do rate automated methods as somewhat fair and are somewhat more willing for more technical evaluation (coding assessments), across the board they always prefer some amount of human involvement to AI-only decision-making. In particular and contrary to the actual implementation of AEDTs, which favor employer priorities, participants showed a preference for the use of automation as long as a human reviewed any rejections. Also, when asked whether they would like to be evaluated using these methods participants' responses were lower than their fairness ratings. On the whole, these results speak to young job seekers' distrust of and distaste for the use of automation in hiring. 

Finally, we also examined participants' strategies for navigating these systems, as well as demographic attributes, and the effect of both on reported job outcomes (whether they were successful in receiving any job offers). Strikingly, we found that most strategies did not significantly impact perceptions or job success. The only predictors of job success were the percentage of jobs applied to with a referral and family income, reflecting the continued role of socioeconomic privilege in the work and in life.
Far from being ``bias-free'' or ``bias-mitigated,'' as many AEDTs self-advertise, our work suggests they may instead exacerbate an existing pattern of inequity, since those more financially or socially privileged may be able to circumvent them altogether. Given these findings, we implore AEDT developers, employers, and policymakers to reconsider the use of entirely automated hiring pipelines (especially for less technical tasks), to proactively consider and combat existing social inequalities, and to create meaningful requirements for transparency and fairness in the use of these tools. 

\bibliographystyle{ACM-Reference-Format}
\bibliography{ref}


\begin{thebibliography}{70}

%%% the Web address.

\ifx \showCODEN    \undefined \def \showCODEN     #1{\unskip}     \fi
\ifx \showDOI      \undefined \def \showDOI       #1{#1}\fi
\ifx \showISBNx    \undefined \def \showISBNx     #1{\unskip}     \fi
\ifx \showISBNxiii \undefined \def \showISBNxiii  #1{\unskip}     \fi
\ifx \showISSN     \undefined \def \showISSN      #1{\unskip}     \fi
\ifx \showLCCN     \undefined \def \showLCCN      #1{\unskip}     \fi
\ifx \shownote     \undefined \def \shownote      #1{#1}          \fi
\ifx \showarticletitle \undefined \def \showarticletitle #1{#1}   \fi
\ifx \showURL      \undefined \def \showURL       {\relax}        \fi
% invisible to TeX
\providecommand\bibfield[2]{#2}
\providecommand\bibinfo[2]{#2}
\providecommand\natexlab[1]{#1}
\providecommand\showeprint[2][]{arXiv:#2}

\bibitem[Adnin et~al\mbox{.}(2022)]%
        {adnin2022hiring}
\bibfield{author}{\bibinfo{person}{Rudaiba Adnin}, \bibinfo{person}{Sadia Afroz}, \bibinfo{person}{Muhtasim Ulfat}, {and} \bibinfo{person}{Anindya Iqbal}.} \bibinfo{year}{2022}\natexlab{}.
\newblock \showarticletitle{A Hiring Story: Experiences of Employers in Hiring CS Graduates in Software Startups}. In \bibinfo{booktitle}{\emph{Companion Publication of the 2022 Conference on Computer Supported Cooperative Work and Social Computing}} (Virtual Event, Taiwan) \emph{(\bibinfo{series}{CSCW'22 Companion})}. \bibinfo{publisher}{Association for Computing Machinery}, \bibinfo{address}{New York, NY, USA}, \bibinfo{pages}{126–129}.
\newblock
\showISBNx{9781450391900}
\urldef\tempurl%
\url{https://doi.org/10.1145/3500868.3559440}
\showDOI{\tempurl}


\bibitem[Ajunwa(2020)]%
        {ajunwa2019paradox}
\bibfield{author}{\bibinfo{person}{Ifeoma Ajunwa}.} \bibinfo{year}{2020}\natexlab{}.
\newblock \showarticletitle{The paradox of automation as anti-bias intervention}.
\newblock \bibinfo{journal}{\emph{Cardozo L. Review}}  \bibinfo{volume}{41} (\bibinfo{year}{2020}), \bibinfo{pages}{1671}.
\newblock
\urldef\tempurl%
\url{http://dx.doi.org/10.2139/ssrn.2746078}
\showURL{%
\tempurl}


\bibitem[Ajunwa(2021)]%
        {ajunwa2019auditing}
\bibfield{author}{\bibinfo{person}{Ifeoma Ajunwa}.} \bibinfo{year}{2021}\natexlab{}.
\newblock \showarticletitle{An Auditing Imperative for Automated Hiring Systems}.
\newblock \bibinfo{journal}{\emph{Harvard Journal of Law and Technology}}  \bibinfo{volume}{34} (\bibinfo{year}{2021}), \bibinfo{pages}{621--699}.
\newblock
\urldef\tempurl%
\url{http://dx.doi.org/10.2139/ssrn.3437631}
\showURL{%
\tempurl}


\bibitem[Armstrong et~al\mbox{.}(2023)]%
        {armstrong2023navigating}
\bibfield{author}{\bibinfo{person}{Lena Armstrong}, \bibinfo{person}{Jayne Everson}, {and} \bibinfo{person}{Amy~J. Ko}.} \bibinfo{year}{2023}\natexlab{}.
\newblock \showarticletitle{Navigating a Black Box: Students’ Experiences and Perceptions of Automated Hiring}. In \bibinfo{booktitle}{\emph{Proceedings of the 2023 ACM Conference on International Computing Education Research - Volume 1}} (Chicago, IL, USA) \emph{(\bibinfo{series}{ICER '23})}. \bibinfo{publisher}{Association for Computing Machinery}, \bibinfo{address}{New York, NY, USA}, \bibinfo{pages}{148–158}.
\newblock
\showISBNx{9781450399760}
\urldef\tempurl%
\url{https://doi.org/10.1145/3568813.3600123}
\showDOI{\tempurl}


\bibitem[Armstrong and Metaxa(2023)]%
        {armstrong2023computing}
\bibfield{author}{\bibinfo{person}{Lena Armstrong} {and} \bibinfo{person}{Dana{\"e} Metaxa}.} \bibinfo{year}{2023}\natexlab{}.
\newblock \showarticletitle{Computing Students’ Strategies and Perceptions of Fairness in Navigating Automated Hiring}.
\newblock  (\bibinfo{year}{2023}).
\newblock
\urldef\tempurl%
\url{https://doi.org/10.17605/OSF.IO/PVM9K}
\showDOI{\tempurl}


\bibitem[Begel and Simon(2008)]%
        {begel2008novice}
\bibfield{author}{\bibinfo{person}{Andrew Begel} {and} \bibinfo{person}{Beth Simon}.} \bibinfo{year}{2008}\natexlab{}.
\newblock \showarticletitle{Novice software developers, all over again}. In \bibinfo{booktitle}{\emph{Proceedings of the Fourth International Workshop on Computing Education Research}} (Sydney, Australia) \emph{(\bibinfo{series}{ICER '08})}. \bibinfo{publisher}{Association for Computing Machinery}, \bibinfo{address}{New York, NY, USA}, \bibinfo{pages}{3–14}.
\newblock
\showISBNx{9781605582160}
\urldef\tempurl%
\url{https://doi.org/10.1145/1404520.1404522}
\showDOI{\tempurl}


\bibitem[Behroozi et~al\mbox{.}(2019)]%
        {behroozi2019hiring}
\bibfield{author}{\bibinfo{person}{Mahnaz Behroozi}, \bibinfo{person}{Chris Parnin}, {and} \bibinfo{person}{Titus Barik}.} \bibinfo{year}{2019}\natexlab{}.
\newblock \showarticletitle{Hiring is Broken: What Do Developers Say About Technical Interviews?}. In \bibinfo{booktitle}{\emph{2019 IEEE Symposium on Visual Languages and Human-Centric Computing (VL/HCC)}}. \bibinfo{pages}{1--9}.
\newblock
\urldef\tempurl%
\url{https://doi.org/10.1109/VLHCC.2019.8818836}
\showDOI{\tempurl}


\bibitem[Bharadwaj et~al\mbox{.}(2022)]%
        {bharadwaj2022resume}
\bibfield{author}{\bibinfo{person}{S Bharadwaj}, \bibinfo{person}{Rudra Varun}, \bibinfo{person}{Potukuchi~Sreeram Aditya}, \bibinfo{person}{Macherla Nikhil}, {and} \bibinfo{person}{G.~Charles Babu}.} \bibinfo{year}{2022}\natexlab{}.
\newblock \showarticletitle{Resume Screening using NLP and LSTM}. In \bibinfo{booktitle}{\emph{2022 International Conference on Inventive Computation Technologies (ICICT)}}. \bibinfo{pages}{238--241}.
\newblock
\urldef\tempurl%
\url{https://doi.org/10.1109/ICICT54344.2022.9850889}
\showDOI{\tempurl}


\bibitem[Binns et~al\mbox{.}(2018)]%
        {binns2018s}
\bibfield{author}{\bibinfo{person}{Reuben Binns}, \bibinfo{person}{Max Van~Kleek}, \bibinfo{person}{Michael Veale}, \bibinfo{person}{Ulrik Lyngs}, \bibinfo{person}{Jun Zhao}, {and} \bibinfo{person}{Nigel Shadbolt}.} \bibinfo{year}{2018}\natexlab{}.
\newblock \showarticletitle{'It's Reducing a Human Being to a Percentage': Perceptions of Justice in Algorithmic Decisions}. In \bibinfo{booktitle}{\emph{Proceedings of the 2018 CHI Conference on Human Factors in Computing Systems}} (Montreal QC, Canada) \emph{(\bibinfo{series}{CHI '18})}. \bibinfo{publisher}{Association for Computing Machinery}, \bibinfo{address}{New York, NY, USA}, \bibinfo{pages}{1–14}.
\newblock
\showISBNx{9781450356206}
\urldef\tempurl%
\url{https://doi.org/10.1145/3173574.3173951}
\showDOI{\tempurl}


\bibitem[Bock et~al\mbox{.}(2013)]%
        {bock2013women}
\bibfield{author}{\bibinfo{person}{Skyler~J Bock}, \bibinfo{person}{Lindsay~J Taylor}, \bibinfo{person}{Zachary~E Phillips}, {and} \bibinfo{person}{Wenying Sun}.} \bibinfo{year}{2013}\natexlab{}.
\newblock \showarticletitle{Women and minorities in computer science majors: Results on barriers from interviews and a survey}.
\newblock \bibinfo{journal}{\emph{Issues in Information Systems}} \bibinfo{volume}{14}, \bibinfo{number}{1} (\bibinfo{year}{2013}), \bibinfo{pages}{143--152}.
\newblock
\urldef\tempurl%
\url{https://iacis.org/iis/2013/189_iis_2013_143-152.pdf}
\showURL{%
\tempurl}


\bibitem[Bogen and Rieke(2018)]%
        {bogen2018help}
\bibfield{author}{\bibinfo{person}{Miranda Bogen} {and} \bibinfo{person}{Aaron Rieke}.} \bibinfo{year}{2018}\natexlab{}.
\newblock \showarticletitle{Help wanted: An examination of hiring algorithms, equity, and bias}.
\newblock \bibinfo{journal}{\emph{Upturn}} (\bibinfo{year}{2018}).
\newblock


\bibitem[Buyl et~al\mbox{.}(2022)]%
        {buyl2022tackling}
\bibfield{author}{\bibinfo{person}{Maarten Buyl}, \bibinfo{person}{Christina Cociancig}, \bibinfo{person}{Cristina Frattone}, {and} \bibinfo{person}{Nele Roekens}.} \bibinfo{year}{2022}\natexlab{}.
\newblock \showarticletitle{Tackling Algorithmic Disability Discrimination in the Hiring Process: An Ethical, Legal and Technical Analysis}. In \bibinfo{booktitle}{\emph{Proceedings of the 2022 ACM Conference on Fairness, Accountability, and Transparency}} (Seoul, Republic of Korea) \emph{(\bibinfo{series}{FAccT '22})}. \bibinfo{publisher}{Association for Computing Machinery}, \bibinfo{address}{New York, NY, USA}, \bibinfo{pages}{1071–1082}.
\newblock
\showISBNx{9781450393522}
\urldef\tempurl%
\url{https://doi.org/10.1145/3531146.3533169}
\showDOI{\tempurl}


\bibitem[Caliskan et~al\mbox{.}(2017)]%
        {caliskan2017semantics}
\bibfield{author}{\bibinfo{person}{Aylin Caliskan}, \bibinfo{person}{Joanna~J. Bryson}, {and} \bibinfo{person}{Arvind Narayanan}.} \bibinfo{year}{2017}\natexlab{}.
\newblock \showarticletitle{Semantics derived automatically from language corpora contain human-like biases}.
\newblock \bibinfo{journal}{\emph{Science}} \bibinfo{volume}{356}, \bibinfo{number}{6334} (\bibinfo{year}{2017}), \bibinfo{pages}{183--186}.
\newblock
\urldef\tempurl%
\url{https://doi.org/10.1126/science.aal4230}
\showDOI{\tempurl}


\bibitem[Chen et~al\mbox{.}(2016)]%
        {chen2016automated}
\bibfield{author}{\bibinfo{person}{Lei Chen}, \bibinfo{person}{Gary Feng}, \bibinfo{person}{Chee~Wee Leong}, \bibinfo{person}{Blair Lehman}, \bibinfo{person}{Michelle Martin-Raugh}, \bibinfo{person}{Harrison Kell}, \bibinfo{person}{Chong~Min Lee}, {and} \bibinfo{person}{Su-Youn Yoon}.} \bibinfo{year}{2016}\natexlab{}.
\newblock \showarticletitle{Automated scoring of interview videos using Doc2Vec multimodal feature extraction paradigm}. In \bibinfo{booktitle}{\emph{Proceedings of the 18th ACM International Conference on Multimodal Interaction}} (Tokyo, Japan) \emph{(\bibinfo{series}{ICMI '16})}. \bibinfo{publisher}{Association for Computing Machinery}, \bibinfo{address}{New York, NY, USA}, \bibinfo{pages}{161–168}.
\newblock
\showISBNx{9781450345569}
\urldef\tempurl%
\url{https://doi.org/10.1145/2993148.2993203}
\showDOI{\tempurl}


\bibitem[Chen et~al\mbox{.}(2018)]%
        {chen2018investigating}
\bibfield{author}{\bibinfo{person}{Le Chen}, \bibinfo{person}{Ruijun Ma}, \bibinfo{person}{Anik{\'o} Hann{\'a}k}, {and} \bibinfo{person}{Christo Wilson}.} \bibinfo{year}{2018}\natexlab{}.
\newblock \showarticletitle{Investigating the {{Impact}} of {{Gender}} on {{Rank}} in {{Resume Search Engines}}}. In \bibinfo{booktitle}{\emph{Proceedings of the 2018 {{CHI Conference}} on {{Human Factors}} in {{Computing Systems}} - {{CHI}} '18}}. \bibinfo{publisher}{{ACM Press}}, \bibinfo{address}{{Montreal QC, Canada}}, \bibinfo{pages}{1--14}.
\newblock
\showISBNx{978-1-4503-5620-6}
\urldef\tempurl%
\url{https://doi.org/10.1145/3173574.3174225}
\showDOI{\tempurl}


\bibitem[Chua et~al\mbox{.}(2021)]%
        {chua2021playing}
\bibfield{author}{\bibinfo{person}{Phoebe~K. Chua}, \bibinfo{person}{Hillary Abraham}, {and} \bibinfo{person}{Melissa Mazmanian}.} \bibinfo{year}{2021}\natexlab{}.
\newblock \showarticletitle{Playing the Hiring Game: Class-Based Emotional Experiences and Tactics in Elite Hiring}.
\newblock \bibinfo{journal}{\emph{Proc. ACM Hum.-Comput. Interact.}} \bibinfo{volume}{5}, \bibinfo{number}{CSCW2}, Article \bibinfo{articleno}{392} (\bibinfo{date}{Oct.} \bibinfo{year}{2021}), \bibinfo{numpages}{27}~pages.
\newblock
\urldef\tempurl%
\url{https://doi.org/10.1145/3479536}
\showDOI{\tempurl}


\bibitem[Colquitt(2012)]%
        {colquitt2012organizational}
\bibfield{author}{\bibinfo{person}{Jason~A. Colquitt}.} \bibinfo{year}{2012}\natexlab{}.
\newblock \showarticletitle{{526 Organizational Justice}}.
\newblock In \bibinfo{booktitle}{\emph{{The Oxford Handbook of Organizational Psychology, Volume 1}}}. \bibinfo{publisher}{Oxford University Press}.
\newblock
\showISBNx{9780199928309}
\urldef\tempurl%
\url{https://doi.org/10.1093/oxfordhb/9780199928309.013.0016}
\showDOI{\tempurl}


\bibitem[Dastin(2018)]%
        {dastin2018amazon}
\bibfield{author}{\bibinfo{person}{Jeffrey Dastin}.} \bibinfo{year}{2018}\natexlab{}.
\newblock \bibinfo{title}{Amazon scraps secret AI recruiting tool that showed bias against women}.
\newblock
\newblock
\urldef\tempurl%
\url{https://www.reuters.com/article/world/insight-amazon-scraps-secret-ai-recruiting-tool-that-showed-bias-against-women-idUSKCN1MK0AG/}
\showURL{%
\tempurl}


\bibitem[De~Cremer and De~Schutter(2021)]%
        {de2021use}
\bibfield{author}{\bibinfo{person}{David De~Cremer} {and} \bibinfo{person}{Leander De~Schutter}.} \bibinfo{year}{2021}\natexlab{}.
\newblock \showarticletitle{How to use algorithmic decision-making to promote inclusiveness in organizations}.
\newblock \bibinfo{journal}{\emph{AI and Ethics}} \bibinfo{volume}{1}, \bibinfo{number}{4} (\bibinfo{year}{2021}), \bibinfo{pages}{563--567}.
\newblock
\urldef\tempurl%
\url{https://doi.org/10.1007/s43681-021-00073-0}
\showDOI{\tempurl}


\bibitem[Dhende et~al\mbox{.}(2018)]%
        {dhende2018candidate}
\bibfield{author}{\bibinfo{person}{Shubham~D Dhende}, \bibinfo{person}{Aniket~S Pashankar}, \bibinfo{person}{Sushil~A Pawar}, \bibinfo{person}{Akash~S Salave}, {and} \bibinfo{person}{Sudhir~D Salunkhe}.} \bibinfo{year}{2018}\natexlab{}.
\newblock \showarticletitle{Candidate hiring through CV analysis}.
\newblock \bibinfo{journal}{\emph{International Research Journal of Engineering and Technology (IRJET)}} \bibinfo{volume}{5}, \bibinfo{number}{05} (\bibinfo{year}{2018}).
\newblock


\bibitem[Dombrowski et~al\mbox{.}(2016)]%
        {dombrowski2016social}
\bibfield{author}{\bibinfo{person}{Lynn Dombrowski}, \bibinfo{person}{Ellie Harmon}, {and} \bibinfo{person}{Sarah Fox}.} \bibinfo{year}{2016}\natexlab{}.
\newblock \showarticletitle{Social Justice-Oriented Interaction Design: Outlining Key Design Strategies and Commitments}. In \bibinfo{booktitle}{\emph{Proceedings of the 2016 ACM Conference on Designing Interactive Systems}} (Brisbane, QLD, Australia) \emph{(\bibinfo{series}{DIS '16})}. \bibinfo{publisher}{Association for Computing Machinery}, \bibinfo{address}{New York, NY, USA}, \bibinfo{pages}{656–671}.
\newblock
\showISBNx{9781450340311}
\urldef\tempurl%
\url{https://doi.org/10.1145/2901790.2901861}
\showDOI{\tempurl}


\bibitem[Farber et~al\mbox{.}(2017)]%
        {farber2017factors}
\bibfield{author}{\bibinfo{person}{Henry~S. Farber}, \bibinfo{person}{Dan Silverman}, {and} \bibinfo{person}{Till~M. von Wachter}.} \bibinfo{year}{2017}\natexlab{}.
\newblock \showarticletitle{Factors Determining Callbacks to Job Applications by the Unemployed: An Audit Study}.
\newblock \bibinfo{journal}{\emph{RSF: The Russell Sage Foundation Journal of the Social Sciences}} \bibinfo{volume}{3}, \bibinfo{number}{3} (\bibinfo{year}{2017}), \bibinfo{pages}{168--201}.
\newblock
\showISSN{2377-8253}
\urldef\tempurl%
\url{https://doi.org/10.7758/RSF.2017.3.3.08}
\showDOI{\tempurl}


\bibitem[Fernandez et~al\mbox{.}(2016)]%
        {fernandez2016more}
\bibfield{author}{\bibinfo{person}{Todd Fernandez}, \bibinfo{person}{Allison Godwin}, \bibinfo{person}{Jacqueline Doyle}, \bibinfo{person}{Dina Verdin}, \bibinfo{person}{Hank Boone}, \bibinfo{person}{Adam Kirn}, \bibinfo{person}{Lisa Benson}, {and} \bibinfo{person}{Geoff Potvin}.} \bibinfo{year}{2016}\natexlab{}.
\newblock \showarticletitle{More comprehensive and inclusive approaches to demographic data collection}.
\newblock  (\bibinfo{year}{2016}).
\newblock


\bibitem[Folger(1977)]%
        {folger1977distributive}
\bibfield{author}{\bibinfo{person}{Robert Folger}.} \bibinfo{year}{1977}\natexlab{}.
\newblock \showarticletitle{Distributive and procedural justice: Combined impact of voice and improvement on experienced inequity.}
\newblock \bibinfo{journal}{\emph{Journal of personality and social psychology}} \bibinfo{volume}{35}, \bibinfo{number}{2} (\bibinfo{year}{1977}), \bibinfo{pages}{108}.
\newblock


\bibitem[Friedrich(1993)]%
        {friedrich1993primary}
\bibfield{author}{\bibinfo{person}{James Friedrich}.} \bibinfo{year}{1993}\natexlab{}.
\newblock \showarticletitle{Primary error detection and minimization (PEDMIN) strategies in social cognition: A reinterpretation of confirmation bias phenomena}.
\newblock \bibinfo{journal}{\emph{Psychological review}} \bibinfo{volume}{100}, \bibinfo{number}{2} (\bibinfo{year}{1993}), \bibinfo{pages}{298}.
\newblock
\urldef\tempurl%
\url{https://doi.org/10.1037/0033-295x.100.2.298}
\showDOI{\tempurl}


\bibitem[Giannakos et~al\mbox{.}(2017)]%
        {giannakos2017understanding}
\bibfield{author}{\bibinfo{person}{Michail~N. Giannakos}, \bibinfo{person}{Ilias~O. Pappas}, \bibinfo{person}{Letizia Jaccheri}, {and} \bibinfo{person}{Demetrios~G. Sampson}.} \bibinfo{year}{2017}\natexlab{}.
\newblock \showarticletitle{Understanding student retention in computer science education: The role of environment, gains, barriers and usefulness}.
\newblock \bibinfo{journal}{\emph{Education and Information Technologies}} \bibinfo{volume}{22}, \bibinfo{number}{5} (\bibinfo{date}{sep} \bibinfo{year}{2017}), \bibinfo{pages}{2365–2382}.
\newblock
\showISSN{1360-2357}
\urldef\tempurl%
\url{https://doi.org/10.1007/s10639-016-9538-1}
\showDOI{\tempurl}


\bibitem[Glikson and Woolley(2020)]%
        {glikson2020human}
\bibfield{author}{\bibinfo{person}{Ella Glikson} {and} \bibinfo{person}{Anita~Williams Woolley}.} \bibinfo{year}{2020}\natexlab{}.
\newblock \showarticletitle{Human trust in artificial intelligence: Review of empirical research}.
\newblock \bibinfo{journal}{\emph{Academy of Management Annals}} \bibinfo{volume}{14}, \bibinfo{number}{2} (\bibinfo{year}{2020}), \bibinfo{pages}{627--660}.
\newblock
\urldef\tempurl%
\url{https://doi.org/10.5465/annals.2018.0057}
\showDOI{\tempurl}


\bibitem[Gonzalez et~al\mbox{.}(2022)]%
        {gonzalez2022allying}
\bibfield{author}{\bibinfo{person}{Manuel~F. Gonzalez}, \bibinfo{person}{Weiwei Liu}, \bibinfo{person}{Lei Shirase}, \bibinfo{person}{David~L. Tomczak}, \bibinfo{person}{Carmen~E. Lobbe}, \bibinfo{person}{Richard Justenhoven}, {and} \bibinfo{person}{Nicholas~R. Martin}.} \bibinfo{year}{2022}\natexlab{}.
\newblock \showarticletitle{Allying with AI? Reactions toward human-based, AI/ML-based, and augmented hiring processes}.
\newblock \bibinfo{journal}{\emph{Comput. Hum. Behav.}} \bibinfo{volume}{130}, \bibinfo{number}{C} (\bibinfo{date}{may} \bibinfo{year}{2022}), \bibinfo{numpages}{16}~pages.
\newblock
\showISSN{0747-5632}
\urldef\tempurl%
\url{https://doi.org/10.1016/j.chb.2022.107179}
\showDOI{\tempurl}


\bibitem[Groves et~al\mbox{.}(2024)]%
        {groves2024auditing}
\bibfield{author}{\bibinfo{person}{Lara Groves}, \bibinfo{person}{Jacob Metcalf}, \bibinfo{person}{Alayna Kennedy}, \bibinfo{person}{Briana Vecchione}, {and} \bibinfo{person}{Andrew Strait}.} \bibinfo{year}{2024}\natexlab{}.
\newblock \showarticletitle{Auditing Work: Exploring the New York City algorithmic bias audit regime}. In \bibinfo{booktitle}{\emph{Proceedings of the 2024 ACM Conference on Fairness, Accountability, and Transparency}} (Rio de Janeiro, Brazil) \emph{(\bibinfo{series}{FAccT '24})}. \bibinfo{publisher}{Association for Computing Machinery}, \bibinfo{address}{New York, NY, USA}, \bibinfo{pages}{1107–1120}.
\newblock
\showISBNx{9798400704505}
\urldef\tempurl%
\url{https://doi.org/10.1145/3630106.3658959}
\showDOI{\tempurl}


\bibitem[Harsha et~al\mbox{.}(2022)]%
        {harsha2022automated}
\bibfield{author}{\bibinfo{person}{Tumula~Mani Harsha}, \bibinfo{person}{Gangaraju~Sai Moukthika}, \bibinfo{person}{Dudipalli~Siva Sai}, \bibinfo{person}{Mannuru Naga~Rajeswari Pravallika}, \bibinfo{person}{Satish Anamalamudi}, {and} \bibinfo{person}{MuraliKrishna Enduri}.} \bibinfo{year}{2022}\natexlab{}.
\newblock \showarticletitle{Automated Resume Screener using Natural Language Processing (NLP)}. In \bibinfo{booktitle}{\emph{2022 6th International Conference on Trends in Electronics and Informatics (ICOEI)}}. \bibinfo{pages}{1772--1777}.
\newblock
\urldef\tempurl%
\url{https://doi.org/10.1109/ICOEI53556.2022.9777194}
\showDOI{\tempurl}


\bibitem[Hickman et~al\mbox{.}(2022)]%
        {hickman2022automated}
\bibfield{author}{\bibinfo{person}{Louis Hickman}, \bibinfo{person}{Nigel Bosch}, \bibinfo{person}{Vincent Ng}, \bibinfo{person}{Rachel Saef}, \bibinfo{person}{Louis Tay}, {and} \bibinfo{person}{Sang~Eun Woo}.} \bibinfo{year}{2022}\natexlab{}.
\newblock \showarticletitle{Automated video interview personality assessments: Reliability, validity, and generalizability investigations.}
\newblock \bibinfo{journal}{\emph{Journal of Applied Psychology}} \bibinfo{volume}{107}, \bibinfo{number}{8} (\bibinfo{year}{2022}), \bibinfo{pages}{1323}.
\newblock


\bibitem[Jebb et~al\mbox{.}(2021)]%
        {jebb2021review}
\bibfield{author}{\bibinfo{person}{Andrew~T Jebb}, \bibinfo{person}{Vincent Ng}, {and} \bibinfo{person}{Louis Tay}.} \bibinfo{year}{2021}\natexlab{}.
\newblock \showarticletitle{A review of key Likert scale development advances: 1995--2019}.
\newblock \bibinfo{journal}{\emph{Frontiers in Psychology}}  \bibinfo{volume}{12} (\bibinfo{year}{2021}), \bibinfo{pages}{637547}.
\newblock
\urldef\tempurl%
\url{https://doi.org/10.3389/fpsyg.2021.637547}
\showURL{%
\tempurl}


\bibitem[Jeung and Huang(2023)]%
        {jeung2023correct}
\bibfield{author}{\bibinfo{person}{Jun~Li Jeung} {and} \bibinfo{person}{Janet Yi-Ching Huang}.} \bibinfo{year}{2023}\natexlab{}.
\newblock \showarticletitle{Correct Me If I Am Wrong: Exploring How AI Outputs Affect User Perception and Trust}. In \bibinfo{booktitle}{\emph{Companion Publication of the 2023 Conference on Computer Supported Cooperative Work and Social Computing}} (Minneapolis, MN, USA) \emph{(\bibinfo{series}{CSCW '23 Companion})}. \bibinfo{publisher}{Association for Computing Machinery}, \bibinfo{address}{New York, NY, USA}, \bibinfo{pages}{323–327}.
\newblock
\showISBNx{9798400701290}
\urldef\tempurl%
\url{https://doi.org/10.1145/3584931.3606997}
\showDOI{\tempurl}


\bibitem[Jones(1986)]%
        {jones1986socialization}
\bibfield{author}{\bibinfo{person}{Gareth~R Jones}.} \bibinfo{year}{1986}\natexlab{}.
\newblock \showarticletitle{Socialization tactics, self-efficacy, and newcomers' adjustments to organizations}.
\newblock \bibinfo{journal}{\emph{Academy of Management journal}} \bibinfo{volume}{29}, \bibinfo{number}{2} (\bibinfo{year}{1986}), \bibinfo{pages}{262--279}.
\newblock


\bibitem[Joshi et~al\mbox{.}(2015)]%
        {joshi2015likert}
\bibfield{author}{\bibinfo{person}{Ankur Joshi}, \bibinfo{person}{Saket Kale}, \bibinfo{person}{Satish Chandel}, {and} \bibinfo{person}{D~Kumar Pal}.} \bibinfo{year}{2015}\natexlab{}.
\newblock \showarticletitle{Likert scale: Explored and explained}.
\newblock \bibinfo{journal}{\emph{British journal of applied science \& technology}} \bibinfo{volume}{7}, \bibinfo{number}{4} (\bibinfo{year}{2015}), \bibinfo{pages}{396--403}.
\newblock
\urldef\tempurl%
\url{https://doi.org/10.9734/BJAST/2015/14975}
\showDOI{\tempurl}


\bibitem[Kasinidou et~al\mbox{.}(2021)]%
        {kasinidou2021agree}
\bibfield{author}{\bibinfo{person}{Maria Kasinidou}, \bibinfo{person}{Styliani Kleanthous}, \bibinfo{person}{P\i{}nar Barlas}, {and} \bibinfo{person}{Jahna Otterbacher}.} \bibinfo{year}{2021}\natexlab{}.
\newblock \showarticletitle{I agree with the decision, but they didn't deserve this: Future Developers' Perception of Fairness in Algorithmic Decisions}. In \bibinfo{booktitle}{\emph{Proceedings of the 2021 ACM Conference on Fairness, Accountability, and Transparency}} (Virtual Event, Canada) \emph{(\bibinfo{series}{FAccT '21})}. \bibinfo{publisher}{Association for Computing Machinery}, \bibinfo{address}{New York, NY, USA}, \bibinfo{pages}{690–700}.
\newblock
\showISBNx{9781450383097}
\urldef\tempurl%
\url{https://doi.org/10.1145/3442188.3445931}
\showDOI{\tempurl}


\bibitem[Lam et~al\mbox{.}(2023)]%
        {lam2023STA}
\bibfield{author}{\bibinfo{person}{Michelle~S. Lam}, \bibinfo{person}{Ayush Pandit}, \bibinfo{person}{Colin~H. Kalicki}, \bibinfo{person}{Rachit Gupta}, \bibinfo{person}{Poonam Sahoo}, {and} \bibinfo{person}{Dana\"{e} Metaxa}.} \bibinfo{year}{2023}\natexlab{}.
\newblock \showarticletitle{Sociotechnical Audits: Broadening the Algorithm Auditing Lens to Investigate Targeted Advertising}.
\newblock \bibinfo{journal}{\emph{Proc. ACM Hum.-Comput. Interact.}} \bibinfo{volume}{7}, \bibinfo{number}{CSCW2}, Article \bibinfo{articleno}{360} (\bibinfo{date}{oct} \bibinfo{year}{2023}), \bibinfo{numpages}{37}~pages.
\newblock
\urldef\tempurl%
\url{https://doi.org/10.1145/3610209}
\showDOI{\tempurl}


\bibitem[Leclercq et~al\mbox{.}(2020)]%
        {leclercq2020gamification}
\bibfield{author}{\bibinfo{person}{Thomas Leclercq}, \bibinfo{person}{Ingrid Poncin}, \bibinfo{person}{Wafa Hammedi}, \bibinfo{person}{Avreliane Kullak}, {and} \bibinfo{person}{Linda~D. Hollebeek}.} \bibinfo{year}{2020}\natexlab{}.
\newblock \showarticletitle{When gamification backfires: the impact of perceived justice on online community contributions}.
\newblock \bibinfo{journal}{\emph{Journal of Marketing Management}} \bibinfo{volume}{36}, \bibinfo{number}{5-6} (\bibinfo{year}{2020}), \bibinfo{pages}{550--577}.
\newblock
\urldef\tempurl%
\url{https://doi.org/10.1080/0267257X.2020.1736604}
\showDOI{\tempurl}


\bibitem[Lee(2018)]%
        {lee2018understanding}
\bibfield{author}{\bibinfo{person}{Min~Kyung Lee}.} \bibinfo{year}{2018}\natexlab{}.
\newblock \showarticletitle{Understanding perception of algorithmic decisions: Fairness, trust, and emotion in response to algorithmic management}.
\newblock \bibinfo{journal}{\emph{Big Data \& Society}} \bibinfo{volume}{5}, \bibinfo{number}{1} (\bibinfo{year}{2018}).
\newblock
\urldef\tempurl%
\url{https://doi.org/10.1177/2053951718756684}
\showDOI{\tempurl}


\bibitem[Lee and Baykal(2017)]%
        {lee2017algorithmic}
\bibfield{author}{\bibinfo{person}{Min~Kyung Lee} {and} \bibinfo{person}{Su Baykal}.} \bibinfo{year}{2017}\natexlab{}.
\newblock \showarticletitle{Algorithmic Mediation in Group Decisions: Fairness Perceptions of Algorithmically Mediated vs. Discussion-Based Social Division}. In \bibinfo{booktitle}{\emph{Proceedings of the 2017 ACM Conference on Computer Supported Cooperative Work and Social Computing}} (Portland, Oregon, USA) \emph{(\bibinfo{series}{CSCW '17})}. \bibinfo{publisher}{Association for Computing Machinery}, \bibinfo{address}{New York, NY, USA}, \bibinfo{pages}{1035–1048}.
\newblock
\showISBNx{9781450343350}
\urldef\tempurl%
\url{https://doi.org/10.1145/2998181.2998230}
\showDOI{\tempurl}


\bibitem[Lee et~al\mbox{.}(2019)]%
        {lee2019procedural}
\bibfield{author}{\bibinfo{person}{Min~Kyung Lee}, \bibinfo{person}{Anuraag Jain}, \bibinfo{person}{Hea~Jin Cha}, \bibinfo{person}{Shashank Ojha}, {and} \bibinfo{person}{Daniel Kusbit}.} \bibinfo{year}{2019}\natexlab{}.
\newblock \showarticletitle{Procedural Justice in Algorithmic Fairness: Leveraging Transparency and Outcome Control for Fair Algorithmic Mediation}.
\newblock \bibinfo{journal}{\emph{Proc. ACM Hum.-Comput. Interact.}} \bibinfo{volume}{3}, \bibinfo{number}{CSCW}, Article \bibinfo{articleno}{182} (\bibinfo{date}{nov} \bibinfo{year}{2019}), \bibinfo{numpages}{26}~pages.
\newblock
\urldef\tempurl%
\url{https://doi.org/10.1145/3359284}
\showDOI{\tempurl}


\bibitem[Leutner et~al\mbox{.}(2023)]%
        {leutner2023game}
\bibfield{author}{\bibinfo{person}{Franziska Leutner}, \bibinfo{person}{Sonia-Cristina Codreanu}, \bibinfo{person}{Suzanne Brink}, {and} \bibinfo{person}{Theodoros Bitsakis}.} \bibinfo{year}{2023}\natexlab{}.
\newblock \showarticletitle{Game based assessments of cognitive ability in recruitment: Validity, fairness and test-taking experience}.
\newblock \bibinfo{journal}{\emph{Frontiers in Psychology}}  \bibinfo{volume}{13} (\bibinfo{year}{2023}).
\newblock
\urldef\tempurl%
\url{https://doi.org/10.3389/fpsyg.2022.942662}
\showDOI{\tempurl}


\bibitem[Leventhal(1980)]%
        {leventhal1980should}
\bibfield{author}{\bibinfo{person}{Gerald~S Leventhal}.} \bibinfo{year}{1980}\natexlab{}.
\newblock \showarticletitle{What should be done with equity theory? New approaches to the study of fairness in social relationships}.
\newblock In \bibinfo{booktitle}{\emph{Social exchange: Advances in theory and research}}. \bibinfo{publisher}{Springer}, \bibinfo{pages}{27--55}.
\newblock


\bibitem[Li et~al\mbox{.}(2021)]%
        {li2021algorithmic}
\bibfield{author}{\bibinfo{person}{Lan Li}, \bibinfo{person}{Tina Lassiter}, \bibinfo{person}{Joohee Oh}, {and} \bibinfo{person}{Min~Kyung Lee}.} \bibinfo{year}{2021}\natexlab{}.
\newblock \showarticletitle{Algorithmic Hiring in Practice: Recruiter and HR Professional's Perspectives on AI Use in Hiring}. In \bibinfo{booktitle}{\emph{Proceedings of the 2021 AAAI/ACM Conference on AI, Ethics, and Society}} (Virtual Event, USA) \emph{(\bibinfo{series}{AIES '21})}. \bibinfo{publisher}{Association for Computing Machinery}, \bibinfo{address}{New York, NY, USA}, \bibinfo{pages}{166–176}.
\newblock
\showISBNx{9781450384735}
\urldef\tempurl%
\url{https://doi.org/10.1145/3461702.3462531}
\showDOI{\tempurl}


\bibitem[Lind et~al\mbox{.}(1990)]%
        {lind1990voice}
\bibfield{author}{\bibinfo{person}{E~Allan Lind}, \bibinfo{person}{Ruth Kanfer}, {and} \bibinfo{person}{P~Christopher Earley}.} \bibinfo{year}{1990}\natexlab{}.
\newblock \showarticletitle{Voice, control, and procedural justice: Instrumental and noninstrumental concerns in fairness judgments.}
\newblock \bibinfo{journal}{\emph{Journal of Personality and Social psychology}} \bibinfo{volume}{59}, \bibinfo{number}{5} (\bibinfo{year}{1990}), \bibinfo{pages}{952}.
\newblock


\bibitem[Metaxa et~al\mbox{.}(2018)]%
        {metaxa2018gender}
\bibfield{author}{\bibinfo{person}{Dana\"{e} Metaxa}, \bibinfo{person}{Kelly Wang}, \bibinfo{person}{James~A. Landay}, {and} \bibinfo{person}{Jeff Hancock}.} \bibinfo{year}{2018}\natexlab{}.
\newblock \showarticletitle{Gender-Inclusive Design: Sense of Belonging and Bias in Web Interfaces}. In \bibinfo{booktitle}{\emph{Proceedings of the 2018 CHI Conference on Human Factors in Computing Systems}} (Montreal QC, Canada) \emph{(\bibinfo{series}{CHI '18})}. \bibinfo{publisher}{Association for Computing Machinery}, \bibinfo{address}{New York, NY, USA}, \bibinfo{pages}{1–6}.
\newblock
\showISBNx{9781450356206}
\urldef\tempurl%
\url{https://doi.org/10.1145/3173574.3174188}
\showDOI{\tempurl}


\bibitem[Morse et~al\mbox{.}(2022)]%
        {morse2021ends}
\bibfield{author}{\bibinfo{person}{Lily Morse}, \bibinfo{person}{Mike Horia~M Teodorescu}, \bibinfo{person}{Yazeed Awwad}, {and} \bibinfo{person}{Gerald~C Kane}.} \bibinfo{year}{2022}\natexlab{}.
\newblock \showarticletitle{Do the ends justify the means? Variation in the distributive and procedural fairness of machine learning algorithms}.
\newblock \bibinfo{journal}{\emph{Journal of Business Ethics}}  \bibinfo{volume}{181} (\bibinfo{year}{2022}), \bibinfo{pages}{1083–1095}.
\newblock
\urldef\tempurl%
\url{https://doi.org/10.1007/s10551-021-04939-5}
\showURL{%
\tempurl}


\bibitem[Naim et~al\mbox{.}(2018)]%
        {naim2016automated}
\bibfield{author}{\bibinfo{person}{Iftekhar Naim}, \bibinfo{person}{Md.~Iftekhar Tanveer}, \bibinfo{person}{Daniel Gildea}, {and} \bibinfo{person}{Mohammed~Ehsan Hoque}.} \bibinfo{year}{2018}\natexlab{}.
\newblock \showarticletitle{Automated Analysis and Prediction of Job Interview Performance}.
\newblock \bibinfo{journal}{\emph{IEEE Transactions on Affective Computing}} \bibinfo{volume}{9}, \bibinfo{number}{2} (\bibinfo{year}{2018}), \bibinfo{pages}{191--204}.
\newblock
\urldef\tempurl%
\url{https://doi.org/10.1109/TAFFC.2016.2614299}
\showDOI{\tempurl}


\bibitem[Nakai and Guo(2023)]%
        {nakai2023uncovering}
\bibfield{author}{\bibinfo{person}{Kendall Nakai} {and} \bibinfo{person}{Philip~J. Guo}.} \bibinfo{year}{2023}\natexlab{}.
\newblock \showarticletitle{Uncovering the Hidden Curriculum of University Computing Majors via Undergraduate-Written Mentoring Guides: A Learner-Centered Design Workflow}. In \bibinfo{booktitle}{\emph{Proceedings of the 2023 ACM Conference on International Computing Education Research - Volume 1}} (Chicago, IL, USA) \emph{(\bibinfo{series}{ICER '23})}. \bibinfo{publisher}{Association for Computing Machinery}, \bibinfo{address}{New York, NY, USA}, \bibinfo{pages}{63–77}.
\newblock
\showISBNx{9781450399760}
\urldef\tempurl%
\url{https://doi.org/10.1145/3568813.3600113}
\showDOI{\tempurl}


\bibitem[of~Consumer and Protection(2023)]%
        {locallaw144}
\bibfield{author}{\bibinfo{person}{NYC~Department of Consumer} {and} \bibinfo{person}{Worker Protection}.} \bibinfo{year}{2023}\natexlab{}.
\newblock \bibinfo{title}{Automated Employment Decision Tools}.
\newblock
\newblock
\urldef\tempurl%
\url{https://codelibrary.amlegal.com/codes/newyorkcity/latest/NYCrules/0-0-0-138391}
\showURL{%
\tempurl}


\bibitem[Olson(2014)]%
        {olson2014opportunities}
\bibfield{author}{\bibinfo{person}{Joann~S. Olson}.} \bibinfo{year}{2014}\natexlab{}.
\newblock \showarticletitle{Opportunities, Obstacles, and Options: First-Generation College Graduates and Social Cognitive Career Theory}.
\newblock \bibinfo{journal}{\emph{Journal of Career Development}} \bibinfo{volume}{41}, \bibinfo{number}{3} (\bibinfo{year}{2014}), \bibinfo{pages}{199--217}.
\newblock
\urldef\tempurl%
\url{https://doi.org/10.1177/0894845313486352}
\showDOI{\tempurl}


\bibitem[Parasurama and Sedoc(2022)]%
        {parasurama2022gendered}
\bibfield{author}{\bibinfo{person}{Prasanna Parasurama} {and} \bibinfo{person}{Jo{\~a}o Sedoc}.} \bibinfo{year}{2022}\natexlab{}.
\newblock \showarticletitle{Gendered information in resumes and its role in algorithmic and human hiring bias}. In \bibinfo{booktitle}{\emph{Academy of Management Proceedings}}, Vol.~\bibinfo{volume}{2022}. Academy of Management Briarcliff Manor, NY 10510, \bibinfo{pages}{17133}.
\newblock
\urldef\tempurl%
\url{https://doi.org/10.5465/AMBPP.2022.285}
\showDOI{\tempurl}


\bibitem[Raghavan et~al\mbox{.}(2020)]%
        {raghavan2020mitigating}
\bibfield{author}{\bibinfo{person}{Manish Raghavan}, \bibinfo{person}{Solon Barocas}, \bibinfo{person}{Jon Kleinberg}, {and} \bibinfo{person}{Karen Levy}.} \bibinfo{year}{2020}\natexlab{}.
\newblock \showarticletitle{Mitigating bias in algorithmic hiring: evaluating claims and practices}. In \bibinfo{booktitle}{\emph{Proceedings of the 2020 Conference on Fairness, Accountability, and Transparency}} (Barcelona, Spain) \emph{(\bibinfo{series}{FAT* '20})}. \bibinfo{publisher}{Association for Computing Machinery}, \bibinfo{address}{New York, NY, USA}, \bibinfo{pages}{469–481}.
\newblock
\showISBNx{9781450369367}
\urldef\tempurl%
\url{https://doi.org/10.1145/3351095.3372828}
\showDOI{\tempurl}


\bibitem[Rhea et~al\mbox{.}(2022)]%
        {rhea2022resume}
\bibfield{author}{\bibinfo{person}{Alene Rhea}, \bibinfo{person}{Kelsey Markey}, \bibinfo{person}{Lauren D'Arinzo}, \bibinfo{person}{Hilke Schellmann}, \bibinfo{person}{Mona Sloane}, \bibinfo{person}{Paul Squires}, {and} \bibinfo{person}{Julia Stoyanovich}.} \bibinfo{year}{2022}\natexlab{}.
\newblock \showarticletitle{Resume Format, LinkedIn URLs and Other Unexpected Influences on AI Personality Prediction in Hiring: Results of an Audit}. In \bibinfo{booktitle}{\emph{Proceedings of the 2022 AAAI/ACM Conference on AI, Ethics, and Society}} (Oxford, United Kingdom) \emph{(\bibinfo{series}{AIES '22})}. \bibinfo{publisher}{Association for Computing Machinery}, \bibinfo{address}{New York, NY, USA}, \bibinfo{pages}{572–587}.
\newblock
\showISBNx{9781450392471}
\urldef\tempurl%
\url{https://doi.org/10.1145/3514094.3534189}
\showDOI{\tempurl}


\bibitem[Roemmich et~al\mbox{.}(2023)]%
        {roemmich2023values}
\bibfield{author}{\bibinfo{person}{Kat Roemmich}, \bibinfo{person}{Tillie Rosenberg}, \bibinfo{person}{Serena Fan}, {and} \bibinfo{person}{Nazanin Andalibi}.} \bibinfo{year}{2023}\natexlab{}.
\newblock \showarticletitle{Values in Emotion Artificial Intelligence Hiring Services: Technosolutions to Organizational Problems}.
\newblock \bibinfo{journal}{\emph{Proc. ACM Hum.-Comput. Interact.}} \bibinfo{volume}{7}, \bibinfo{number}{CSCW1}, Article \bibinfo{articleno}{109} (\bibinfo{date}{apr} \bibinfo{year}{2023}), \bibinfo{numpages}{28}~pages.
\newblock
\urldef\tempurl%
\url{https://doi.org/10.1145/3579543}
\showDOI{\tempurl}


\bibitem[Roulin et~al\mbox{.}(2014)]%
        {roulin2014interviewers}
\bibfield{author}{\bibinfo{person}{Nicolas Roulin}, \bibinfo{person}{Adrian Bangerter}, {and} \bibinfo{person}{Julia Levashina}.} \bibinfo{year}{2014}\natexlab{}.
\newblock \showarticletitle{Interviewers’ perceptions of impression management in employment interviews}.
\newblock \bibinfo{journal}{\emph{Journal of Managerial Psychology}}  \bibinfo{volume}{29} (\bibinfo{date}{03} \bibinfo{year}{2014}), \bibinfo{pages}{141--163}.
\newblock
\urldef\tempurl%
\url{https://doi.org/10.1108/JMP-10-2012-0295}
\showDOI{\tempurl}


\bibitem[S\'{a}nchez-Monedero et~al\mbox{.}(2020)]%
        {sanchez2020does}
\bibfield{author}{\bibinfo{person}{Javier S\'{a}nchez-Monedero}, \bibinfo{person}{Lina Dencik}, {and} \bibinfo{person}{Lilian Edwards}.} \bibinfo{year}{2020}\natexlab{}.
\newblock \showarticletitle{What does it mean to 'solve' the problem of discrimination in hiring? social, technical and legal perspectives from the UK on automated hiring systems}. In \bibinfo{booktitle}{\emph{Proceedings of the 2020 Conference on Fairness, Accountability, and Transparency}} (Barcelona, Spain) \emph{(\bibinfo{series}{FAT* '20})}. \bibinfo{publisher}{Association for Computing Machinery}, \bibinfo{address}{New York, NY, USA}, \bibinfo{pages}{458–468}.
\newblock
\showISBNx{9781450369367}
\urldef\tempurl%
\url{https://doi.org/10.1145/3351095.3372849}
\showDOI{\tempurl}


\bibitem[Sanyal et~al\mbox{.}(2017)]%
        {sanyal2017resume}
\bibfield{author}{\bibinfo{person}{Satyaki Sanyal}, \bibinfo{person}{Souvik Hazra}, \bibinfo{person}{Soumyashree Adhikary}, {and} \bibinfo{person}{Neelanjan Ghosh}.} \bibinfo{year}{2017}\natexlab{}.
\newblock \showarticletitle{Resume parser with natural language processing}.
\newblock \bibinfo{journal}{\emph{International Journal of Engineering Science}}  \bibinfo{volume}{4484} (\bibinfo{year}{2017}).
\newblock
\urldef\tempurl%
\url{https://doi.org/10.13140/RG.2.2.11709.05607}
\showDOI{\tempurl}


\bibitem[Satheesh et~al\mbox{.}(2020)]%
        {satheesh2020resume}
\bibfield{author}{\bibinfo{person}{K Satheesh}, \bibinfo{person}{A Jahnavi}, \bibinfo{person}{L Iswarya}, \bibinfo{person}{K Ayesha}, \bibinfo{person}{G Bhanusekhar}, {and} \bibinfo{person}{K Hanisha}.} \bibinfo{year}{2020}\natexlab{}.
\newblock \showarticletitle{Resume Ranking based on Job Description using SpaCy NER model}.
\newblock \bibinfo{journal}{\emph{International Research Journal of Engineering and Technology}} \bibinfo{volume}{7}, \bibinfo{number}{05} (\bibinfo{year}{2020}), \bibinfo{pages}{74--77}.
\newblock


\bibitem[Schoeffer and Kuehl(2021)]%
        {schoeffer2021appropriate}
\bibfield{author}{\bibinfo{person}{Jakob Schoeffer} {and} \bibinfo{person}{Niklas Kuehl}.} \bibinfo{year}{2021}\natexlab{}.
\newblock \showarticletitle{Appropriate Fairness Perceptions? On the Effectiveness of Explanations in Enabling People to Assess the Fairness of Automated Decision Systems}. In \bibinfo{booktitle}{\emph{Companion Publication of the 2021 Conference on Computer Supported Cooperative Work and Social Computing}} (Virtual Event, USA) \emph{(\bibinfo{series}{CSCW '21 Companion})}. \bibinfo{publisher}{Association for Computing Machinery}, \bibinfo{address}{New York, NY, USA}, \bibinfo{pages}{153–157}.
\newblock
\showISBNx{9781450384797}
\urldef\tempurl%
\url{https://doi.org/10.1145/3462204.3481742}
\showDOI{\tempurl}


\bibitem[Schumann et~al\mbox{.}(2020)]%
        {schumann2020we}
\bibfield{author}{\bibinfo{person}{Candice Schumann}, \bibinfo{person}{Jeffrey~S. Foster}, \bibinfo{person}{Nicholas Mattei}, {and} \bibinfo{person}{John~P. Dickerson}.} \bibinfo{year}{2020}\natexlab{}.
\newblock \showarticletitle{We Need Fairness and Explainability in Algorithmic Hiring}. In \bibinfo{booktitle}{\emph{Proceedings of the 19th International Conference on Autonomous Agents and MultiAgent Systems}} (Auckland, New Zealand) \emph{(\bibinfo{series}{AAMAS '20})}. \bibinfo{publisher}{International Foundation for Autonomous Agents and Multiagent Systems}, \bibinfo{address}{Richland, SC}, \bibinfo{pages}{1716–1720}.
\newblock
\showISBNx{9781450375184}
\urldef\tempurl%
\url{https://dl.acm.org/doi/abs/10.5555/3398761.3398960}
\showURL{%
\tempurl}


\bibitem[Sloane et~al\mbox{.}(2022)]%
        {sloane2022silicon}
\bibfield{author}{\bibinfo{person}{Mona Sloane}, \bibinfo{person}{Emanuel Moss}, {and} \bibinfo{person}{Rumman Chowdhury}.} \bibinfo{year}{2022}\natexlab{}.
\newblock \showarticletitle{A Silicon Valley love triangle: Hiring algorithms, pseudo-science, and the quest for auditability}.
\newblock \bibinfo{journal}{\emph{Patterns}} \bibinfo{volume}{3}, \bibinfo{number}{2} (\bibinfo{year}{2022}).
\newblock
\urldef\tempurl%
\url{https://doi.org/10.1016/j.patter.2021.100425}
\showURL{%
\tempurl}


\bibitem[Thibaut and Walker(1975)]%
        {thibaut1975procedural}
\bibfield{author}{\bibinfo{person}{John~W Thibaut} {and} \bibinfo{person}{Laurens Walker}.} \bibinfo{year}{1975}\natexlab{}.
\newblock \bibinfo{booktitle}{\emph{Procedural justice: A psychological analysis}}.
\newblock \bibinfo{publisher}{L. Erlbaum Associates}.
\newblock


\bibitem[Tyler(2003)]%
        {tyler2003procedural}
\bibfield{author}{\bibinfo{person}{Tom~R Tyler}.} \bibinfo{year}{2003}\natexlab{}.
\newblock \showarticletitle{Procedural justice, legitimacy, and the effective rule of law}.
\newblock \bibinfo{journal}{\emph{Crime and justice}}  \bibinfo{volume}{30} (\bibinfo{year}{2003}), \bibinfo{pages}{283--357}.
\newblock


\bibitem[van Esch et~al\mbox{.}(2021)]%
        {van2021job}
\bibfield{author}{\bibinfo{person}{Patrick van Esch}, \bibinfo{person}{J~Stewart Black}, {and} \bibinfo{person}{Denni Arli}.} \bibinfo{year}{2021}\natexlab{}.
\newblock \showarticletitle{Job candidates’ reactions to AI-enabled job application processes}.
\newblock \bibinfo{journal}{\emph{AI and Ethics}}  \bibinfo{volume}{1} (\bibinfo{year}{2021}), \bibinfo{pages}{119--130}.
\newblock
\urldef\tempurl%
\url{https://doi.org/10.1007/s43681-020-00025-0}
\showDOI{\tempurl}


\bibitem[Wang et~al\mbox{.}(2020)]%
        {wang2020factors}
\bibfield{author}{\bibinfo{person}{Ruotong Wang}, \bibinfo{person}{F.~Maxwell Harper}, {and} \bibinfo{person}{Haiyi Zhu}.} \bibinfo{year}{2020}\natexlab{}.
\newblock \showarticletitle{Factors Influencing Perceived Fairness in Algorithmic Decision-Making: Algorithm Outcomes, Development Procedures, and Individual Differences}. In \bibinfo{booktitle}{\emph{Proceedings of the 2020 CHI Conference on Human Factors in Computing Systems}} (Honolulu, HI, USA) \emph{(\bibinfo{series}{CHI '20})}. \bibinfo{publisher}{Association for Computing Machinery}, \bibinfo{address}{New York, NY, USA}, \bibinfo{pages}{1–14}.
\newblock
\showISBNx{9781450367080}
\urldef\tempurl%
\url{https://doi.org/10.1145/3313831.3376813}
\showDOI{\tempurl}


\bibitem[Wilson et~al\mbox{.}(2021)]%
        {wilson2021building}
\bibfield{author}{\bibinfo{person}{Christo Wilson}, \bibinfo{person}{Avijit Ghosh}, \bibinfo{person}{Shan Jiang}, \bibinfo{person}{Alan Mislove}, \bibinfo{person}{Lewis Baker}, \bibinfo{person}{Janelle Szary}, \bibinfo{person}{Kelly Trindel}, {and} \bibinfo{person}{Frida Polli}.} \bibinfo{year}{2021}\natexlab{}.
\newblock \showarticletitle{Building and Auditing Fair Algorithms: A Case Study in Candidate Screening}. In \bibinfo{booktitle}{\emph{Proceedings of the 2021 ACM Conference on Fairness, Accountability, and Transparency}} (Virtual Event, Canada) \emph{(\bibinfo{series}{FAccT '21})}. \bibinfo{publisher}{Association for Computing Machinery}, \bibinfo{address}{New York, NY, USA}, \bibinfo{pages}{666–677}.
\newblock
\showISBNx{9781450383097}
\urldef\tempurl%
\url{https://doi.org/10.1145/3442188.3445928}
\showDOI{\tempurl}


\bibitem[Woodruff et~al\mbox{.}(2018)]%
        {woodruff2018qualitative}
\bibfield{author}{\bibinfo{person}{Allison Woodruff}, \bibinfo{person}{Sarah~E. Fox}, \bibinfo{person}{Steven Rousso-Schindler}, {and} \bibinfo{person}{Jeffrey Warshaw}.} \bibinfo{year}{2018}\natexlab{}.
\newblock \showarticletitle{A Qualitative Exploration of Perceptions of Algorithmic Fairness}. In \bibinfo{booktitle}{\emph{Proceedings of the 2018 CHI Conference on Human Factors in Computing Systems}} (Montreal QC, Canada) \emph{(\bibinfo{series}{CHI '18})}. \bibinfo{publisher}{Association for Computing Machinery}, \bibinfo{address}{New York, NY, USA}, \bibinfo{pages}{1–14}.
\newblock
\showISBNx{9781450356206}
\urldef\tempurl%
\url{https://doi.org/10.1145/3173574.3174230}
\showDOI{\tempurl}


\bibitem[Wright et~al\mbox{.}(2024)]%
        {wright2024null}
\bibfield{author}{\bibinfo{person}{Lucas Wright}, \bibinfo{person}{Roxana~Mika Muenster}, \bibinfo{person}{Briana Vecchione}, \bibinfo{person}{Tianyao Qu}, \bibinfo{person}{Pika~(Senhuang) Cai}, \bibinfo{person}{Alan Smith}, \bibinfo{person}{Comm 2450~Student Investigators}, \bibinfo{person}{Jacob Metcalf}, {and} \bibinfo{person}{J.~Nathan Matias}.} \bibinfo{year}{2024}\natexlab{}.
\newblock \showarticletitle{Null Compliance: NYC Local Law 144 and the challenges of algorithm accountability}. In \bibinfo{booktitle}{\emph{Proceedings of the 2024 ACM Conference on Fairness, Accountability, and Transparency}} (Rio de Janeiro, Brazil) \emph{(\bibinfo{series}{FAccT '24})}. \bibinfo{publisher}{Association for Computing Machinery}, \bibinfo{address}{New York, NY, USA}, \bibinfo{pages}{1701–1713}.
\newblock
\showISBNx{9798400704505}
\urldef\tempurl%
\url{https://doi.org/10.1145/3630106.3658998}
\showDOI{\tempurl}


\bibitem[Zhang and Yencha(2022)]%
        {zhang2022examining}
\bibfield{author}{\bibinfo{person}{Lixuan Zhang} {and} \bibinfo{person}{Christopher Yencha}.} \bibinfo{year}{2022}\natexlab{}.
\newblock \showarticletitle{Examining perceptions towards hiring algorithms}.
\newblock \bibinfo{journal}{\emph{Technology in Society}}  \bibinfo{volume}{68} (\bibinfo{year}{2022}), \bibinfo{pages}{101848}.
\newblock
\showISSN{0160-791X}
\urldef\tempurl%
\url{https://doi.org/10.1016/j.techsoc.2021.101848}
\showDOI{\tempurl}


\end{thebibliography}

\appendix
\section{Appendix}
\label{appendix}

\subsection{Survey Questions}
\label{app:survey}

\subsubsection{Scenarios}

Participants were asked about three classes of hiring scenarios: technical coding assessments, resume review, and behavioral interviews (the scenarios are listed by class below). For each scenario, they answered two questions, both on 5-point Likert scales:
\begin{itemize}
    \item How fair does this hiring process seem to you? (``This hiring process seems fair'', 1: Strongly disagree to 5: Strongly agree)
    \item If you were applying for a technology job, would you want to be evaluated this way? (``I want to be evaluated this way'', 1: Strongly disagree to 5: Strongly agree)
\end{itemize}

[Technical Coding Assessments]
\begin{enumerate}
\item An applicant submits a sample of code, which is reviewed by a recruitment team, who determines whether the applicant advances to the next phase.
\item An applicant is given an online coding assessment, which is evaluated by an algorithm. If the applicant reaches a certain score on the autograder, the applicant advances to the next phase. All algorithmic decisions are reviewed by a recruitment team.
\item An applicant is given an online coding assessment, which is evaluated by an algorithm. If the algorithm rejects the applicant, the decision is reviewed by a recruitment team. 
\item An applicant is given an online coding assessment, which is evaluated by an algorithm. If the algorithm advances the applicant to the next phase, the decision is reviewed by a recruitment team. 
\item An applicant is given an online coding assessment, which is evaluated by an algorithm that determines whether an applicant advances to the next phase. 
% \item Why did you select the answers above for the different scenarios related to coding assessments?
\end{enumerate}

[Resume Review]
\begin{enumerate}
\item An applicant submits a resume, which is reviewed by a recruitment team, who determines whether the applicant advances to the next phase.
\item An applicant submits a resume, which is evaluated by an algorithm. The algorithm determines whether the applicant advances to the next phase. All algorithmic decisions are reviewed by a recruitment team. 
\item An applicant submits a resume, which is evaluated by an algorithm. If the algorithm rejects your application, the decision is reviewed by a recruitment team. 
\item An applicant submits a resume, which is evaluated by an algorithm. If the algorithm advances the applicant to the next phase, the decision is reviewed by a recruitment team. 
\item An applicant submits a resume, which is evaluated by an algorithm that determines whether an applicant advances to the next phase. 
% \item Why did you select the answers above for the different scenarios related to resumes?
\end{enumerate}

[Behavioral Interviews]
\begin{enumerate}
\item An applicant has an interview with a member of the recruitment team. The recruitment team determines whether the applicant advances to the next phase.
\item An applicant participates in an automated video interview, where the applicant receives interview questions and records video responses. The video, including the applicant’s speech (fluency, prosody, pronunciation, language usage) and nonverbal behaviors (facial expressions, posture, and eye movements), is evaluated by an algorithm. Whether you advance to the next phase is determined by the algorithm. All algorithmic decisions are reviewed by a recruitment team.
\item An applicant participates in an automated video interview, where the applicant receives interview questions and records video responses. The video, including the applicant’s speech (fluency, prosody, pronunciation, language usage) and nonverbal behaviors (facial expressions, posture, and eye movements), is evaluated by an algorithm. If the algorithm rejects the applicant,  the decision is reviewed by a recruitment team. 
\item An applicant participates in an automated video interview, where the applicant receives interview questions and records video responses. The video, including the applicant’s speech (fluency, prosody, pronunciation, language usage) and nonverbal behaviors (facial expressions, posture, and eye movements), is evaluated by an algorithm. If the algorithm advances the applicant to the next phase, the decision is reviewed by a recruitment team. 
\item An applicant participates in an automated video interview, where the applicant receives interview questions and records video responses. The video, including the applicant’s speech (fluency, prosody, pronunciation, language usage) and nonverbal behaviors (facial expressions, posture, and eye movements), is evaluated by an algorithm that determines whether an applicant advances to the next phase.
% \item Why did you select the answers above for the different scenarios related to interviews?
\end{enumerate}

At the end of each set of Likert questions, participants were also asked an open response question (``Why did you select the answers above for the different scenarios related to [coding assessments/resumes/interviews]?'').

\subsubsection{Awareness of AEDTs}

In this section, participants were asked for each hiring process (online coding assessment, automated resume readers, and automated interviews) to check the box to indicate whether they have experience or knowledge of it:
\begin{itemize}
    \item[$\square$] Yes, I have experienced it
    \item[$\square$] No, but I have heard of it
    \item[$\square$] I'm not sure, but have heard of it
    \item[$\square$] No, I have not heard of or experienced it
\end{itemize}

Participants also responded to ``I know how my data was used in the hiring process'' and ``I received feedback from automated hiring algorithms'' from 1: Strongly disagree to 5: Strongly agree.

\subsubsection{Strategy Use}

Participants were asked the following questions about strategy use:
\begin{itemize}
\item Have you modified your resume specifically for automated resume readers? (Yes/No)
\item Have you added keywords from your job description? (Yes/No)
\item Have you changed the layout? (Yes/No)
\item Have you put it through a resume scanner? (Yes/No)
\item Have you modified your resume in some other way for automated hiring? (please specify)
\item Did you use a tool (LeetCode, HackerRank, etc.) to practice for coding assessments? (Yes/No)
\item Have you used anything else to prepare for automated assessments? (please specify)
\item Have you ever received a job referral? (Yes/No)
\item What proportion of your job applications did you have a referral for? (approximate percentage)
\item Approximately how many companies did you apply to? 
\item How did you learn about the application process? (check all that apply)
    \begin{itemize}
        \item[$\square$] Application materials and descriptions
        \item[$\square$] Online resources
        \item[$\square$] Career services through university 
        \item[$\square$] People who had gone through the application process
        \item[$\square$] Recruiter outside of company
        \item[$\square$] Recruiter through company
        \item[$\square$] Family members who worked at companies 
        \item[$\square$] Friends who worked at companies 
        \item[$\square$] Other people who worked at companies
    \end{itemize}
There was also an option to include additional strategies and an attention check in this stage.
\end{itemize}

\subsubsection{Hiring Outcome}
Participants were also asked about their hiring process and its outcome.
\begin{itemize}
\item Have you completed your hiring process? (Yes/No/Not applying to jobs)
\item I am satisfied with my hiring process so far. (1: Strongly disagree to 5: Strongly agree)
\item What is the outcome of your hiring process so far? 
    \begin{itemize}
        \item[$\square$] Multiple job offers
        \item[$\square$] One job offer
        \item[$\square$] No job offers
        \item[$\square$] Not applying to jobs
    \end{itemize}
\end{itemize}

\subsubsection{Demographic Information}
All questions in this section were optional and asked participants to disclose demographic information.

\begin{itemize}
    \item How would you describe your gender identity? (Select all that apply)
        \begin{itemize}
            \item[$\square$] Woman
            \item[$\square$] Man
            \item[$\square$] Non-binary
            \item[$\square$] Genderqueer
            \item[$\square$] Agender
            \item[$\square$] A gender not listed
        \end{itemize}
    \item What best describes you? (Select all that apply)
        \begin{itemize}
            \item[$\square$] Black or African-American
            \item[$\square$] American Indian or Alaskan Native
            \item[$\square$] Asian American or Asian
            \item[$\square$] Hispanic or Latinx
            \item[$\square$] Middle Eastern or North African
            \item[$\square$] Pacific Islander
            \item[$\square$] White or Caucasian
            \item[$\square$] Some other race, ethnicity, or origin 
        \end{itemize}
    \item What is your family’s approximate household income? 
\end{itemize}

\clearpage 

\subsection{Complete Statistical Results}
\label{app:stats}

\begin{table}[ht]
\begin{tabular}{lrrrrl}
\hline
\textbf{}                                            & \textbf{Estimate} & \textbf{Std. Error} & \textbf{t value} & \textbf{Pr(\textgreater{}|t|)} & \textbf{} \\ \hline
(Intercept)                                       & 2.786  & 0.266 & 10.493 & \textless{}0.01 &   \\
Added job description keywords to resume & 0.139  & -1.468    & 0.144 & 0.121            &   \\
Modified resume layout for resume readers & 0.150         & 0.133           & 1.119            & 0.265                         &           \\
Put resume through a resume scanner               & 0.001  & 0.136 & 0.007  & 0.995           &   \\
Practiced for online coding assessment            & 0.249  & 0.140 & 1.787  & 0.075           &   \\
Used referrals                                    & -0.336 & 0.136 & -2.478 & 0.014           & * \\
Percent of companies applied to with referral   & 0.002         & 0.003           & 0.817            & 0.415                         &           \\
Number of companies applied to                    & 0.001  & -0.516    & 0.606 & 0.405            &   \\
Awareness of online coding assessments            & -0.551 & 0.235 & -2.349 & 0.020           & * \\
Awareness of resume scanners                      & 0.014  & 0.183 & 0.076  & 0.940           &   \\
Awareness of automated video interviews           & 0.354  & 0.170 & 2.113  & 0.036           & * \\
Knowledge of data use                             & 0.055  & 0.047 & 1.162  & 0.247           &   \\
Received feedback in the hiring process           & 0.058  & 0.046 & 1.257  & 0.210           &   \\
Used application materials and descriptions       & -0.176 & 0.114 & -1.539 & 0.125           &   \\
Used online resources                             & 0.288  & 0.133 & 2.160  & 0.032           & * \\
Used career services through university           & 0.063  & 0.108 & 0.588  & 0.557           &   \\
Talked with people who had recently applied       & 0.129  & 0.127 & 1.012  & 0.313           &   \\
Connected with recruiter outside of company       & 0.053  & 0.159 & 0.336  & 0.737           &   \\
Connected with recruiter through company          & 0.124  & -1.346    & 0.180 & 0.191            &   \\
Had family who worked at companies        & 0.044  & 0.144 & 0.306  & 0.760           &   \\
Had friends who worked at companies               & 0.140  & 0.112 & 1.247  & 0.214           &   \\
Connected with other company contacts             & -0.022 & 0.126 & -0.179 & 0.858           &   \\
Race                                              & 0.005  & 0.109 & 0.425  & 0.671           &   \\
Gender                                            & -0.003 & 0.142 & -0.024 & 0.981           &   \\
Income                                            & 0.0000002  & 0.0000003 & 0.569  & 0.570           &   \\ \hline
\end{tabular}
\caption{\label{tab:fairStats} Linear regression model of procedural fairness perceptions for automated processes based on strategy use, awareness of AEDTs, gender, race, and income.}
\end{table}

\begin{table}[ht]
\begin{tabular}{lrrrrl}
\hline
\textbf{}                                            & \textbf{Estimate} & \textbf{Std. Error} & \textbf{t value} & \textbf{Pr(\textgreater{}|t|)} & \textbf{} \\ \hline
(Intercept)                                 & 2.479  & 0.268 & 9.267  & \textless{}0.01 &    \\
Added job description keywords to resume    & 0.140         & -1.374              & 0.171           & 0.210                         &           \\
Modified resume layout for resume readers & 0.169         & 0.135          & 1.257            & 0.210                         &           \\
Put resume through a resume scanner         & 0.038  & 0.137 & 0.273  & 0.785           &    \\
Practiced for online coding assessment      & 0.201  & 0.141 & 1.427  & 0.155           &    \\
Used referrals                              & -0.316 & 0.137 & -2.312 & 0.022           & *  \\
Percent of companies applied to with referral   & 0.002         & 0.003           & 0.670            & 0.504                         &           \\
Number of companies applied to              & 0.0004  & 0.001 & 0.544  & 0.589           &    \\
Awareness of online coding assessments      & -0.557 & 0.237 & -2.356 & 0.019           & *  \\
Awareness of resume scanners                & -0.046 & 0.184 & -0.248 & 0.805           &    \\
Awareness of automated video interviews     & 0.440  & 0.169 & 2.608  & {0.010}           & * \\
Knowledge of data use                       & 0.106  & 0.047 & 2.240  & 0.026           & *  \\
Received feedback in the hiring process     & 0.027  & 0.046 & 0.588  & 0.558           &    \\
Used application materials and descriptions & -0.220 & 0.012 & -1.911 & 0.057           &    \\
Used online resources                       & 0.261  & 0.134 & 1.942  & 0.054           &    \\
Used career services through university     & 0.152  & 0.108 & 1.399  & 0.163           &    \\
Talked with people who had recently applied & 0.172  & 0.128 & 1.344  & 0.181           &    \\
Connected with recruiter outside of company & 0.160  & -0.005    & 0.996 & 0.180          &    \\
Connected with recruiter through company    & 0.125  & -1.392    & 0.165 & 0.968           &    \\
Had family who worked at companies  & -0.006 & 0.145 & -0.040 & 0.968           &    \\
Had friends who worked at companies         & 0.134  & 0.113 & 1.188  & 0.236           &    \\
Connected with other company contacts       & 0.049  & 0.127 & 0.385  & 0.700           &    \\
Race                                        & 0.013  & 0.110 & 0.122  & 0.903           &    \\
Gender                                      & -0.116 & 0.143 & -0.815 & 0.416           &    \\
Income                                      & 0.0000002  & 0.0000003 & 0.623  & 0.534           &    \\ \hline
\end{tabular}
\caption{\label{tab:evalStats} Linear regression model of willingness to be evaluated by automated processes based on strategy use, awareness of AEDTs, gender, race, and income.}
\end{table}

\clearpage

\begin{table}[ht]
\begin{tabular}{lrrrrrl}
\toprule
& \textbf{Estimate}  & \textbf{Std. Error} & \textbf{t value} & \textbf{Pr(\textgreater{}|t|)} &   \\
\hline
(Intercept)                                          & 0.329     & 0.237      & 1.386   & 0.168                &   \\
Added job description keywords to resume    & 0.168     & 0.107      & 1.563   & 0.121                 &   \\
Modified resume layout for resume readers & 0.103     & -0.724     & 0.471   & 0.515                 &   \\
Put resume through a resume scanner                  & 0.020     & 0.101      & 0.201   & 0.841                 &   \\
Practiced for online coding assessment               & -0.201    & 0.133      & -1.513  & 0.133                 &   \\
Used referrals                                       & 0.122     & 0.100      & 1.213   & 0.227                 &   \\
Percent of companies applied to with referral   & 0.004     & 0.002      & 2.063   & 0.041                 & * \\
Number of companies applied to                       & 0.0004    & 0.001     & 0.835   & 0.405                 &   \\
Awareness of online coding assessments               & 0.050     & 0.199      & 0.252   & 0.801                 &   \\
Awareness of resume scanners                         & -0.019    & 0.173      & -0.109  & 0.913                 &   \\
Awareness of automated video interviews              & -0.036    & 0.157      & -0.228  & 0.820                 &   \\
Knowledge of data use                                & 0.039     & 0.004      & 0.984   & 0.327                 &   \\
Received feedback in the hiring process              & 0.011     & 0.004      & 0.302   & 0.763                 &   \\
Used application materials and descriptions          & 0.025     & 0.009      & 0.279   & 0.781                 &   \\
Used online resources                                & -0.174    & 0.115      & -1.518  & 0.132                 &   \\
Used career services through university              & 0.055     & 0.085      & 0.644   & 0.521                 &   \\
Talked with people who had recently applied          & 0.024     & 0.107      & 0.225   & 0.823                 &   \\
Connected with recruiter outside of company          & 0.009     & 0.112      & 0.080   & 0.937                 &   \\
Connected with recruiter through company             & 0.115     & 0.088      & 1.314   & 0.191                 &   \\
Had family who worked at companies           & -0.140    & 0.109      & -1.287  & 0.200                 &   \\
Had friends who worked at companies                  & 0.160     & 0.087      & 1.841   & 0.068                 &   \\
Connected with other company contacts                & -0.101    & 0.093     & -1.089  & 0.278                &   \\
Race                                                 & -0.008    & 0.119      & -0.070  & 0.945                 &   \\
Gender                                               & 0.081     & 0.081     & 0.991   & 0.324                 &   \\
Income                                               & 0.000001 & 0.0000002  & 2.530   & 0.013                 & * \\
\bottomrule
\end{tabular}
\caption{\label{tab:jobStats} Linear regression model of job success based on strategy use, awareness of AEDTs, gender, race, and income.}
\end{table}

\end{document}